%% file: 0-main.tex
\documentclass[sigconf,nonacm]{acmart}
\makeatletter
\@ACM@balancefalse
\makeatother

\input{00-packages.tex}
\input{00-commands.tex}
\input{00-acm_commands.tex}

\begin{document}
\title{Mining Path Association Rules in Large Property Graphs\\ (with Appendix)}

\input{00-authors.tex}
\input{00-abstract.tex}


\maketitle

\input{01-introduction}
\input{02-relatedwork}
\input{03-concept}
\input{04-algorithm}
\input{05-approx}
\input{06-parallel}
\input{07-experiment}
\input{08-conclusion}

\section*{Acknowledge}
This work was supported by Japan Science and Technology Agency (JST) as part of Adopting Sustainable Partnerships for Innovative Research Ecosystem (ASPIRE), Grant Number JPMJAP2328, and JST Presto Grant Number JPMJPR21C5. P. Karras was supported by the Independent Research Fund Denmark (Project~9041-00382B).

\bibliographystyle{ACM-Reference-Format}
\bibliography{graphassociationrule}

\appendix
\input{90-appendix}

\end{document}

%% file: 00-packages.tex
\usepackage[ruled, vlined, linesnumbered]{algorithm2e}
\usepackage{algorithmic}
\usepackage{epsfig}
\usepackage{xspace}
\usepackage{booktabs,subcaption,amsfonts,dcolumn}
\usepackage{multirow}
\usepackage{xfrac} 
\usepackage{multicol} 

%% file: 00-commands.tex
\newcommand{\graph}{\ensuremath{\mathcal{G}}}
\newcommand{\edges}{\ensuremath{\mathcal{E}}}
\newcommand{\vertices}{\ensuremath{\mathcal{V}}}
\newcommand{\elabel}{\ensuremath{\ell}}
\newcommand{\elabels}{\ensuremath{\mathcal{L}}}
\newcommand{\attributes}{\ensuremath{\mathcal{A}}}
\newcommand{\rules}{\ensuremath{\mathcal{R}}}

\newcommand{\pathpatterns}{\ensuremath{\mathcal{P}}}
\newcommand{\minsup}{\ensuremath{\theta}}
\newcommand{\minlen}{\ensuremath{k}}
\newcommand{\candidates}{\ensuremath{\mathcal{C}}}
\newcommand{\labelfont}[1]{\texttt{#1}}

\newcommand{\pk}[1]{{#1}} 
\newcommand{\pkk}[1]{{{#1}}} 

\newcommand{\name}{PARM\xspace} 

\newcommand{\method}{{\sc Pioneer}\xspace} 

\newtheorem{example}{Example}
\newtheorem{definition}{Definition}
\newtheorem{problem}{Problem}
\newtheorem{theorem}{Theorem}
\newtheorem{lemma}{Lemma}

%% file: 00-acm_commands.tex
\AtBeginDocument{%
  }

\setcopyright{acmcopyright}
\copyrightyear{2023}
\acmYear{2023}
\acmDOI{XXXXXXX.XXXXXXX}

\acmConference[Conference acronym 'XX]{Make sure to enter the correct
  conference title from your rights confirmation email}{June 03--05,
  2023}{Woodstock, NY}
\acmPrice{15.00}
\acmISBN{978-1-4503-XXXX-X/18/06}

%% file: 00-authors.tex
\author{Yuya Sasaki}
\affiliation{%
  \institution{Osaka University}
  \country{Japan}
}
\email{sasaki@ist.osaka-u.ac.jp}

\author{Panagiotis Karras}
\affiliation{%
  \institution{University of Copenhagen}
  \country{Denmark}
}
\email{piekarras@gmail.com}

%% file: 00-abstract.tex
\begin{abstract}
How can we mine frequent path regularities from a graph with edge labels and vertex attributes? The task of association rule mining successfully discovers regular patterns in item sets and substructures. Still, to our best knowledge, this concept has not yet been extended to path patterns in large property graphs. In this paper, we introduce the problem of \emph{path association rule mining} (\name). Applied to any \emph{reachability path} between two vertices within a large graph, \name discovers regular ways in which path patterns, identified by vertex attributes and edge labels, co-occur with each other. 
We develop an efficient and scalable algorithm \method that exploits an anti-monotonicity property to effectively prune the search space. Further, we devise approximation techniques and employ parallelization to achieve scalable path association rule mining. Our experimental study using real-world graph data verifies the significance of path association rules and the efficiency of our solutions.
\end{abstract}

%% file: 01-introduction.tex
\section{Introduction}\label{sec:introduction}

{\bf Association rule mining} is the task of discovering regular correlation patterns among data objects in large data collections~\cite{agrawal1993mining, zhao2003association}. An association rule, represented as~$X \Rightarrow Y$, where~$X$ is an \emph{antecedent} set, list, or other structure and~$Y$ is a corresponding~\emph{consequent}, signifies that a data record containing~$X$ is likely to also contain~$Y$. Association rules are useful in applications ranging from web mining~\cite{lee2001web} to market analysis~\cite{kaur2016market} and bioinformatics~\cite{mallik2014ranwar}.

\smallskip\noindent {\bf Graph association rule mining} aims to discover regularities among entities on a single large graph~\cite{fan2015association, wang2020extending, fan2022discovering}. A graph association rule is represented as~$G_X \Rightarrow G_Y$, where~$G_X$ and~$G_Y$ are graph patterns. Since graphs are widely used in many applications, the mining of association rules from graphs promises to discover valuable insights and knowledge. Its applications include:

\noindent \underline{\it{Social analysis}}: Graph association rule mining can be used to discover regularities in social relationships. For example, as social relationship patterns affect people's health~\cite{house1988social} and happiness~\cite{haller2006social}, we may discover a rule like ``\pkk{\emph{people who identify as happy are likely to connect with others who also identify as happy through multiple intermediaries with high probability}.}''

\noindent \underline{\it{Discrimination checking}}: Machine learning models trained on graph data are vulnerable to discriminatory bias~\cite{fisher2019measuring}. For example, automated systems reviewing applicant resumes incorporated a significant bias in favor of male candidates due to bias inherent in the training data~\cite{bias2018}. To build fair machine learning models, we should eschew such data-driven discrimination. Since graph association rules discover regularities, they can reveal discriminatory bias.

\noindent \underline{\it{Knowledge extraction}}: Knowledge bases are often represented as graphs with labeled edges and attributed vertices, known as \emph{knowledge graphs}~\cite{weikum21}. We can mine interesting patterns from such graphs as association rules. For example, an interesting rule may be ``people often have occupations similar to those of some of their ancestors.''

\begin{figure*}[t]
\centering
\includegraphics[width=1.0\linewidth]{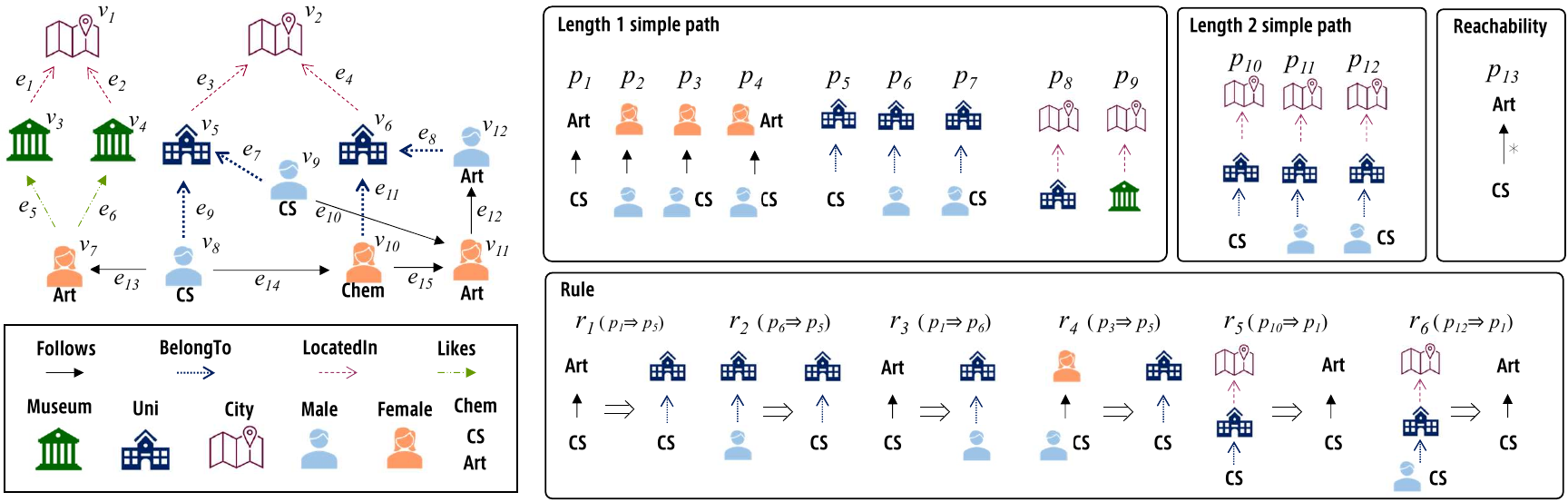}
\vspace{-6mm}
\caption{Path association rule mining.  }\label{fig:intro}
\vspace{-2mm}
\end{figure*}

\smallskip\noindent {\bf Motivation.} While graph association rule mining on a single large graph is fundamental for graph analysis, existing methods~\cite{fan2015association, wang2020extending, fan2022discovering} are inapplicable to the aforementioned applications due to the following shortcomings: (i) regarding the \emph{vertices} in~$G_Y$ and~$G_X$,  existing methods~\cite{fan2015association, fan2022discovering} require the vertex set in the consequent~$G_Y$ to be a subset of that in the antecedent~$G_X$ and mainly focus on missing edges --- association rules where~$G_Y$ includes vertices not in~$G_X$ is out of their scope; (ii) they consider specific restricted graph patterns, e.g., a single edge in~$G_Y$~\cite{fan2015association}, a subgraph including at least one edge in~$G_X$ and~$G_Y$~\cite{wang2020extending}, or a subgraph without attributes in~$G_X$ and a single edge or attribute in~$G_Y$~\cite{fan2022discovering}, and cannot handle edge labels and vertex attributes together; and (iii) their antecedent and consequent do not capture \emph{reachability} (or transitive closure) patterns, which denote that one vertex is reachable by any number of label-constrained directed edges from another, such as the one regarding the examples on \pkk{social analysis and knowledge extraction. Therefore, we need a different approach to graph association rule mining that addresses these shortcomings and thereby ensure wide applicability.}

\smallskip\noindent {\bf Contribution.} In this paper, we introduce a novel, simple, and elegant concept, \emph{path association rule}, which expresses regular co-occurrences of sequences of vertex attribute sets and edge labels, or \emph{path patterns}, and allow for measures such as \emph{absolute/relative support}, \emph{confidence}, and \emph{lift}. Such rules are in the form~$p_X \Rightarrow p_Y$, where~$p_X$ and~$p_Y$ are path patterns with a minimum support of~$\minsup$ common source vertices. To mine path association rules while eschewing the aforementioned shortcomings, we propose a novel, efficient, and scalable algorithm that imposes no restriction on how the vertices in the consequent relate to those in the antecedent, accommodates reachability patterns, and considers both vertex attributes and edge labels.

\vspace{-1mm}
\begin{example}
Figure~\ref{fig:intro} presents an example of path association rule mining on a social network with 12 vertices, 15 edges,
4 types of edge labels $\{ \labelfont{Follows}, \labelfont{BelongTo},$ $\labelfont{LocatedIn},$ $\labelfont{Likes} \}$ and 8 types of vertex attributes $\{ \labelfont{Museum}, \labelfont{Uni}, \labelfont{City}, \labelfont{Male}, \labelfont{Female},$ $\labelfont{CS}, \labelfont{Chem}, \labelfont{Art} \}$. 
We set~$\minsup = 2$; then path pattern $p_1=\langle \{ \labelfont{CS} \}, \labelfont{Follows}, \{\labelfont{Art} \} \rangle$ is frequent, as it has source vertices (or \emph{matches})~$v_8$ and~$v_9$; likewise, path pattern~$p_6 = \langle \{\labelfont{Male} \}, \labelfont{BelongTo}, \{\labelfont{Uni} \} \rangle$ matches~$v_8$ and~$v_9$. We mine path association rule $r_3=\langle \{\labelfont{CS} \}, \labelfont{Follows}, \{\labelfont{Art} \} \rangle$ $\Rightarrow$ $\langle \{\labelfont{Male} \}, \\\labelfont{BelongTo},$ $\{\labelfont{Uni} \} \rangle$, i.e., frequently a computer scientist who follows an artist is male and attends a university. This rule indicates biases between gender/major and education.
Existing methods cannot mine this rule, since the set of vertices in~$G_Y$ is not a subset of those in~$G_X$.
\end{example}
\vspace{-1mm}

We \pkk{develop a novel algorithm, \method (Path associatION rulE minER), which exploits the anti-monotonicity property of path association rules to prune candidate} frequent path patterns and hence rules. We prune candidates by two means: \emph{vertical pruning} stops extending the length of path patterns; \emph{horizontal} pruning stops extending the attributes in path patterns. Our algorithm computes exact support and confidence measures. Further, we develop a probabilistic approximation scheme and a parallelization technique to render path association rule mining more scalable.
It achieves efficient rule mining even for a computationally intractable problem.

Our extensive experiments on four real graphs demonstrate that our exact algorithm accelerates path association rule mining up to~151 times over a baseline, while our approximation scheme is up to~485 times faster with a small accuracy loss. We also show that PARM is effective in checking discrimination and extracting knowledge. Overall, our contributions are summarized as follows:

\begin{itemize}
    \item {\bf Concept}: We propose path association rule mining, \name, which captures regularities between path patterns in a single large graph and manages reachability patterns, unlike existing graph association rule mining.
    \item {\bf Algorithm}: We develop an efficient and scalable algorithm \method that mines rules in parallel while pruning infrequent path patterns and admits an approximation scheme.
    \item {\bf Discoveries}: We show that path association rule mining is effective in bias checking and knowledge extraction.
\end{itemize}

\noindent {\bf Reproducibility}. Our codebase is available.\footnote{\url{https://github.com/yuya-s/pathassociationrulemining}} All proofs are in the appendix.

%% file: 02-relatedwork.tex
\input{tables/relatedwork.tex}
\section{Related Work}\label{sec:related}

We review existing works and research topics related to our work.

\noindent
{\bf Frequent graph mining on a single large graph}.
Definitions of frequency (i.e., support) in a single graph are different across studies. 
The support measures applied on transaction data do not preserve anti-monotonicity properties on a single graph. That is because, intuitively, the number of paths in a graph is usually larger than the number of vertices, even though paths are more complex than vertices. Several support measures that enforce anti-monotonicity properties have been proposed, such as maximum independent set based support (MIS)~\cite{vanetik2002computing} minimum-image-based support (MNI)~\cite{bringmann2008frequent}, minimum clique partition (MCP)~\cite{calders2008anti}, minimum vertex cover (MVC)~\cite{meng2017flexible}, and maximum independent edge set support (MIES)~\cite{meng2017flexible}. Their common goal is to use anti-monotonic properties in case a vertex is involved in multiple matches. However, these support measures have three drawbacks. First, they do not apply to \emph{relative} support because it is hard to count the maximum number of graph patterns that may appear in a graph; while the support measure proposed in~\cite{fan2015association} can be applied to relative support for a single large graph, it is inefficient because it needs isomorphic subgraph matching. 
Second, their time complexity is very high. For instance, the problems of computing MIS and MNI are NP-hard. Third, they are not intuitive, as it is difficult to understand why some vertices match graph patterns and others do not.

Each algorithm on frequent subgraph mining in a single graph uses anti-monotonic properties specialized for their support. Existing supports for subgraph mining did not consider how to handle reachability patterns. 
Support of subgraphs with reachability patterns is untrivial, and thus, it is hard to directly apply algorithms for frequent subgraph mining to our problem.

Their basic concepts of frequent subgraph mining algorithms consist of (1) finding small sizes of frequent patterns, (2) combining frequent patterns to generate new candidates of patterns, (3) removing infrequent patterns based on anti-monotonic properties, and (4) repeating until candidates are empty. Commonly, steps (2) and (3) are extended to efficient processing for their patterns. 
Our baseline in Sec.~\ref{sec:baseline} follows this basic method. 

\noindent
{\bf Graph pattern mining}.
Several algorithms have been developed for graph pattern mining~
\cite{elseidy2014grami, shelokar2014three, deng2021mining, fariha2013mining, prateek2020mining, ke2009efficient, nikolakaki2018mining, alipourlangouri2022discovery}, 
each with different semantics. For example, Prateek et al.~\cite{prateek2020mining} introduce a method for finding pairs of subgraphs that appear often in close proximity; Nikolakaki et al.~\cite{nikolakaki2018mining} propose a method that finds a set of diverse paths that minimize a cost of overlapping edges and vertices. However, graph pattern mining does not handle reachability path patterns. On the other hand, algorithms for isomorphic subgraph matching, e.g., \cite{han2019efficient, fan2020extending, lee2012depth} aim to efficiently discover matching patterns in a single large graph. These methods are not suitable for frequent pattern mining, as they need to find each different subgraph pattern from scratch.

\noindent
{\bf Graph association rule mining}.
Graph association rule mining applies to two type of data: a set of (transactional) graphs and a single large graph. Methods for transactional graph data and those for a single large graph are not interchangeable because their anti-monotonicity properties are different. Algorithms for a set of graphs aim to find rules that apply in at least~\minsup{} graphs in the collection (e.g., \cite{inokuchi2000apriori,ke2009efficient,yan2002gspan,samiullah2014mining}).
On the other hand, algorithms for a single large graph aim to find rules that appear in a single graph at least~\minsup{} times~\cite{fan2017big, fan2015association, wang2020extending, fan2022discovering,namaki2017discovering,huynh2022mining}. To our best knowledge, none of these methods focuses on paths or reachability patterns. Table~\ref{table:comparison} shows the characteristics of such methods, including ours.



\emph{Graph pattern association rules} on a large single graph, or GPARs, were introduced in~\cite{fan2015association}. Their association rules focus on specific patterns where the consequent specifies a single edge and a set of vertices that is a subset of the vertices in antecedent. A rule evaluates whether the antecedent contains the edge specified in the consequent. Besides, the algorithm in~\cite{fan2015association} aims to find rules with a fixed consequent rather than all valid frequent rules, so it is hard to extend to the latter direction. They use a vertex-centric support measure that counts the number of vertices in a subgraph that match a specified pivot; this measure allows for measuring relative support via extensive subgraph matching.

Certain works extend or generalize GPAR. Wang et al.~\cite{wang2020extending} find association rules using the MIS support measure~\cite{vanetik2002computing} and require the antecedent and consequent to be subgraphs with at least one edge each but no common edges. Yet this technique cannot use relative support and cannot find regularities among vertex attributes (e.g., occupation and gender) because it does not allow specifying a single property as consequent. Fan et al.~\cite{fan2022discovering} proposed \emph{graph association rules}, or GARs, that generalize GPARs with vertex attributes; this is the only graph association rule mining method that handles \emph{both} edge labels and vertex attributes, albeit it allows only a \emph{single} edge or attribute in the consequent. It also provides sampling to reduce graph size according to a set of required graph patterns in~$G_Y$. The difference between this sampling method and ours is that our sampling reduces the candidate source vertices, while the GAR algorithm reduces the graph itself.

The GPAR is applicable to find missing edge patterns in \emph{quantified} graph patterns~\cite{fan2016adding,fan2017big} that include potential and quantified edges and to discover temporal regularities on dynamic graphs~\cite{namaki2017discovering}.

\noindent
{\bf Mining other rules}.
Several studies extract other rules from graphs, such as graph functional dependencies~\cite{fan2020discovering} and Horn rules~\cite{manola2004rdf, galarraga2013amie, chen2016scalekb, meilicke2020reinforced, ortona2018robust, chen2022rule}, which are similar to a path association rules. In a Horn rule, a consequent is a single edge whose vertices are included in the antecedent on RDF data. 
Yet Horn rules do not cover general property graphs.

Subsequence mining~\cite{agrawal1995mining, zaki1998planmine, nowozin2007discriminative, fournier2015mining} finds regularities of sub-sequence patterns in sequences. These are a special type of graph association rules, since a sequence can be seen as a path graph. Yet subsequence mining methods cannot apply to complex graphs.

%% file: tables/relatedwork.tex
\begin{table*}[t!]
\caption{Methods for association rule mining on a single large graph. $V.G_X$ and $V.G_Y$ are the vertex sets of~$G_X$ and~$G_Y$, respectively.}\label{table:comparison_duplicate}
\vspace{-3mm}
\scalebox{0.85}{
\begin{tabular}{l|lllll}
\hline
                                        &Graph type& $G_X$ & $G_Y$ & $G_X$ and $G_Y$ &Output\\ \hline
\multirow{2}{*}{GPAR~\cite{fan2015association}}         &labeled edges& \multirow{2}{*}{Subgraph} & Single edge & \multirow{2}{*}{$V. G_Y \subseteq V.G_X$} & \multirow{2}{*}{Top-$k$ diverse patterns} \\
                                        &labeled vertices &  & or empty & & \\ \hline
\multirow{2}{*}{Extending GPAR~\cite{wang2020extending}} &unlabeled edges& Subgraph & Subgraph  & $G_Y$ is connected to $G_X$ & \multirow{2}{*}{Frequent patterns}\\
                                        & attributed vertices & (at least one edge) &  (at least one edge) & and no common edges  & \\\hline
\multirow{2}{*}{GAR~\cite{fan2022discovering}}           &labeled edges& Subgraph & Single edge & \multirow{2}{*}{$V. G_X \subseteq V.G_Y$} & \multirow{2}{*}{Application-specific frequent pattern}\\
 &attributed vertices  &without attributes  & or single attribute & & \\ \hline
\multirow{2}{*}{PARM (Ours)}                                  &labeled edges& \multirow{2}{*}{Path} &\multirow{2}{*}{Path} & \multirow{2}{*}{Sources are common} & \multirow{2}{*}{Frequent patterns}\\
 &attributed vertices  &  & & & \\ 
\hline
\end{tabular}}
\end{table*}

%% file: 03-concept.tex
\section{The Concept}\label{sec:concept}

\pk{We propose the novel concept of \emph{path association rule mining} (PARM), which effectively discovers regularities among attributes of vertices connected by labeled edges. The distinctive characteristic of PARM compared to existing graph association rule mining techniques is that it captures correlations of distinct path patterns among the same vertices, which are useful in many applications. In addition, PARM discovers rules on general property graphs, which cover many graph types (e.g., labeled graphs).}

\subsection{Notations}

\pk{We consider a graph~$\graph=(\vertices,\edges,\elabels,\attributes)$, where~$\vertices$ is a set of vertices, $\edges \subset \vertices \times \elabels \times \vertices$ is a set of edges, $\elabels$ is a set of edge labels, and~$\attributes$ is a set of attributes. Each edge~$e \in \edges$ is a triple $(v, \elabel_e, v')$ denoting an edge from vertex~$v$ to vertex~$v'$ with label~$\elabel_e$. Attribute~$a \in \attributes$ is a categorical value representing a feature of a vertex. Each vertex~$v \in \vertices$ has a set of attributes~$A(v) \subseteq \attributes$.}

\pk{A \emph{path} is a sequence of vertices and edges~$\langle v_0, e_0, v_1, \ldots, e_{n-1}, v_n \rangle$, where~$n$ is its \emph{length}, $v_0$ its \emph{source}, and~$v_n$ its \emph{target}. A path \emph{prefix} (\emph{suffix}) is an arbitrary initial (final) part of a path.}

\begin{example}
\pk{In Figure~\ref{fig:intro}, $\elabels =\{\labelfont{Follows}, \labelfont{BelongTo},$ $\labelfont{LocatedIn},$ $\labelfont{Likes} \}$ and $\attributes = \{\labelfont{Museum}, \labelfont{Uni}, \labelfont{City}, \labelfont{Male}, \labelfont{Female},$ $\labelfont{CS}, \labelfont{Chem}, \labelfont{Art} \}$. $\langle v_8, $ $e_9, v_5, e_3, v_2 \rangle$ is a path of length~$2$ path with source~$v_8$ and target~$v_2$.}
\end{example}

\subsection{Path Association Rules}

\pk{We define path association rules after defining path patterns.

\noindent
{\bf Path pattern}. We define simple and reachability path patterns.

\begin{itemize}
\item A \emph{simple path pattern} is a sequence of attribute sets and edge labels~$\langle A_0, \elabel_0, A_1, \ldots, \elabel_{n-1}, A_n \rangle$ where~$A_i \!\subseteq \!\attributes$ ($A_i \neq \varnothing$) and~$\elabel_i \!\in\! \elabels$; $n$ indicates the pattern's length.

\item A \emph{reachability path pattern} is a pair of attribute sets with an edge label~$\langle A_0, \elabel^*, A_1 \rangle$, where~$A_0, A_1  \subseteq \attributes$ ($A_0, A_1 \neq \varnothing$) and~$\elabel^* \in \elabels$.
\end{itemize}

We say that a path~$\langle v_0, e_0, v_1, \ldots, e_{n-1}, v_n \rangle$ matches a simple path pattern~$\langle A_0, \elabel_0, A_1, \ldots,$ $\elabel_{n-1}, A_n \rangle$ if~$A_i \!\! \subseteq \!\! A(v_i)$ and~$\elabel_i \!\! = \!\!\elabel_{e_i}$ for all~$i$. Similarly, a path matches a reachability path pattern~$\langle A_0, \elabel^*, A_1 \rangle$ if~$A_0 \subseteq A(v_0)$, $A_1 \subseteq A(v_n)$, and~$\elabel^* \!\! = \!\!\elabel_{e_i}$ for all $i$.

Given a vertex~$v$ and a path pattern~$p$, $v$ \emph{matches}~$p$ if it is the \emph{source} of a path that matches~$p$. $\vertices(p)$ denotes the set of all vertices matching path pattern~$p$ and~$|\vertices(p)|$ its cardinality. Given a positive integer~$\minsup$, we say that~$p$ is frequent if~$|\vertices(p)| \geq \minsup$. A \emph{unit path pattern} is a path pattern such that each of its attribute sets comprises a single attribute.}

\begin{definition}[Dominance]
\pk{Given two path patterns~$p = \langle A_0, \elabel_0,$ $\ldots, \elabel_{n-1}, A_n \rangle$ and~$p'=\langle A_0',$ $\elabel_0',$ $\ldots, \elabel_{m-1}', A_m' \rangle$, we say that~$p$ \emph{dominates}~$p'$ if~$m \leq n$, $\elabel_i' = \elabel_i$, and~$A_i' \subseteq A_i$ for~$i = 0$ to~$m$.}
\end{definition}

\pk{ $p' \subset p$ indicates $p$ dominates $p'$. Intuitively, a dominating path pattern is more complex than its dominated path patterns.}

\begin{example}
\pk{In Figure~\ref{fig:intro}, $v_9$ matches simple path patterns $\langle \{\labelfont{CS},$ $\labelfont{Male} \}, \labelfont{Follows}, \{\labelfont{Art}, \labelfont{Female} \} \rangle$ and reachability path patterns $\langle \{\labelfont{CS}$, $\labelfont{Male} \},$ $\labelfont{Follows}^{*}, \{\labelfont{Art}, \labelfont{Male} \} \rangle$. The vertex set~$\vertices(\langle \{\labelfont{Male}\},$ $\labelfont{Follows},$ $\{\labelfont{Female}\} \rangle)$ is~$\{v_8,$ $v_9 \}$.}
\end{example}

\noindent
\pk{{\bf Path association rule}. We define path association rules as follows.

\vspace{-1mm}
\begin{definition}
A path association rule~$r$ is expressed as~$p_X \Rightarrow p_Y$, where~$p_X$ and~$p_Y$ are path patterns; $p_X$ is the antecedent and~$p_Y$ is the consequent of the rule. We say that a vertex matches~$r$ if it matches both~$p_X$ and~$p_Y$.
\end{definition}

Given a path association rule, we may evaluate the frequency and conditional probability of vertices that match both~$p_X$ and~$p_Y$. We apply {\it homomorphism semantics}, allowing a single path to be shared by both~$p_X$ and~$p_Y$.}


\begin{example}
\pk{The rule~$\langle \{\labelfont{CS} \}, \labelfont{Follows}, \{\labelfont{Art}\} \rangle \Rightarrow$ $\langle \{\labelfont{CS} \},$ $\labelfont{BelongTo}$, $\{\labelfont{Uni}\} \rangle$ in Figure~\ref{fig:intro} matches~$v_8$ and~$v_9$.}
\end{example}


\subsection{Measures of association rules}

Path association rules support measures similar to those of association rules~\cite{agrawal1993mining}. Here, we define support, confidence, and lift for path association rules.

\noindent\underline{\it{Support}}:
The \emph{support} of a path association rule~$r$ indicates how many vertices it matches.
We define \emph{absolute} and \emph{relative} support. Significantly, most graph association rule mining methods do not offer relative support (see Section~\ref{sec:related}), as it is hard to compute the number of matched graph patterns. Absolute support is defined as $\mathit{ASupp}(p_X \Rightarrow p_Y) = |\vertices(p_X) \cap \vertices(p_Y)|$. Since the maximum value of~$|\vertices(p)|$ is the number of vertices, relative support is defined as $\mathit{RSupp}(p_X\Rightarrow p_Y) = \frac{|\vertices(p_X) \cap \vertices(p_Y)|}{|\vertices|}$. Relative measures assess effectiveness independently of data size~\cite{karras08}.

\noindent
\underline{\it{Confidence}}:
The \emph{confidence} of a path association rule indicates the probability that a vertex satisfies~$p_Y$ if given it satisfies~$p_X$. We define confidence as $\mathit{Conf}(p_X \Rightarrow p_Y) = \frac{|\vertices(p_X) \cap \vertices(p_Y)|}{|\vertices(p_X)|}$.

\noindent
\underline{\it{Lift}}: Lift, which most graph association rule mining methods do not support, quantifies how much the probability of~$p_Y$ is lifted given the antecedent~$p_X$, compared to its unconditioned counterpart. We define lift as $\mathit{Lift}(p_X \Rightarrow p_Y) = \frac{ |\vertices(p_X) \cap \vertices(p_Y)|\cdot |\vertices|}{|\vertices(p_X)|\cdot|\vertices(p_Y)|}$.


\begin{example}
\pk{In Figure~\ref{fig:intro}, for~$r_1 = p_1 \Rightarrow p_5$, we have~$\vertices(p_1) = \vertices(p_5) = \{ v_8, v_9\}$. Since~$|\vertices| = 12$ and~$|\vertices(p_1) \cap \vertices(p_5)| = 2$, it is~$\mathit{ASupp(r_1)} = 2$, $\mathit{RSupp(r_1)} = \sfrac{1}{6}$, $\mathit{Conf(r_1)} = 1$, and~$\mathit{Lift(r_1)} = 6$.}
\end{example}

\subsection{Problem Definition}

\pk{We now define the problem that we solve in this paper.

\begin{problem}[Path Association Rule Mining (\name)]
Given graph~\graph, minimum (absolute or relative) support~\minsup, and maximum path length~\minlen, the path association rule mining problem calls to find all rules~$r$, where (1)~$\mathit{supp}(r) \geq \minsup$, (2)~path lengths are at most \minlen, and (3)~$p_X$ is not dominated by~$p_Y$ or vice versa.
\end{problem}

\smallskip\noindent
{\bf Remark}. Path association rule mining generalizes conventional association rule mining for itemsets~\cite{agrawal1993mining}. We may consider each vertex as a transaction and its attributes as items in the transaction. Ignoring edges, path association rule mining degenerates to conventional itemset-based association rule mining.}

\smallskip\noindent
{\bf Extensions}. We can modify \name according to the application, for example, we can find rules where~$p_Y$ dominates~$p_X$, a set of attributes in path patterns is empty, and we can specify other thresholds (e.g., high confidence and Jaccard similarity) to restrict the number of outputs. We may also use other quality measures (e.g., \cite{pellegrina2019hypothesis, de2013subjective}) in place of the conventional measures we employ. As we focus on introducing the \name problem, we relegate such extensions and investigations to future work.



%% file: 04-algorithm.tex
\section{Main \name Algorithm}\label{sec:algorithm}

We now present our core algorithm for efficient path association rule mining. To solve the \name problem, we need to (1) enumerate frequent path patterns as candidates,
(2) find vertices matching frequent path patterns, and
(3) derive rules.
Our \name algorithm first finds simple frequent path patterns and then generates more complex path pattern candidates therefrom using anti-monotonicity properties. Its efficiency is based on effectively containing the number of candidates.



\input{04.1.antimonoton}
\input{04.2.baseline}
\input{04.3.optim}
\input{04.4.auxiliary}
\input{04.5.pseudocode}

%% file: 04.1.antimonoton.tex
\subsection{Anti-monotonicity properties}\label{sec:anti}

\pk{The following anti-monotonicity properties facilitate the reduction of candidate path patterns.}

\begin{theorem}\label{theo:dominate}
If $p \subset p'$, then $\vertices(p) \supseteq \vertices(p')$.
\end{theorem}


\noindent
\pk{The following two lemmata follow from Theorem~\ref{theo:dominate}.}

\begin{lemma}\label{lemma:antimonotonicity_1}
\pk{If the prefix of a path pattern is infrequent, the whole pattern is infrequent.}
\end{lemma}


\begin{lemma}\label{lemma:antimonotonicity_2}
\pk{If path pattern~$p$ is infrequent and~$p'$ comprises~$A_i' \supseteq A_i$ and~$\elabel_i' = \elabel_i$ for all~$i$, then~$p'$ is also infrequent.}
\end{lemma}


\begin{example}
\pk{We illustrate these anti-monotonicity properties using Figure~\ref{fig:intro}. Let the minimum absolute support be~$2$. Path pattern $p = \langle \{ \labelfont{CS} \}, \labelfont{Follows}, \{\labelfont{Chem} \} \rangle$ is infrequent, thus no path pattern that extends~$p$ is frequent either. Similarly, path patterns that dominate~$p$, such as~$\langle \{ \labelfont{Male,CS} \}, \labelfont{Follows}, \{\labelfont{Female,Chem} \} \rangle$, are infrequent.}
\end{example}

%% file: 04.2.baseline.tex
\subsection{Baseline algorithm}\label{sec:baseline}

Our baseline algorithm discovers, in a sequence, (1) frequent attribute sets, (2) frequent simple path patterns, (3) frequent reachability path patterns, and (4) rules by Lemma~\ref{lemma:antimonotonicity_1}.
We describe each step in the following.

\begin{enumerate}
\item {\bf Frequent attribute set discovery.} We obtain a set $\pathpatterns_0$ of frequent attribute sets, i.e., path patterns of length~0. For this step, we employ any algorithm for conventional association rule mining, such as Apriori~\cite{agrawal1993mining}.

\item {\bf Frequent simple path pattern discovery.} We iteratively extend the discovered frequent path patterns to derive frequent simple path patterns. To generate a candidate simple path pattern of length~$i$, we combine~$p \in \pathpatterns_{i-1}$, an edge label, and~$A \subseteq \attributes$; we then check the pattern's frequency and, if it is frequent, we store it as a frequent simple path pattern in~$\pathpatterns_{i}$. We repeat until the path pattern length becomes \minlen. \label{step:simple}

\item {\bf  Frequent reachability path pattern discovery.} To find vertices that match reachability path patterns, we find, for each vertex~$v$, the set of vertices that are reached through~$\elabel^*$ from~$v$. For each edge label, we enumerate reachable vertices from vertices that have frequent attributes by breadth-first search. Then, to generate a candidate reachability path pattern, we combine~$p \in \pathpatterns_{0}$, an edge label, and~$A \subseteq \attributes$; we then count the number of its matched vertices, and, if it is frequent, store it in~$\pathpatterns^*$. \label{step:reachability}

\item {\bf Rule discovery.} To discover rules, we search for vertices that match two path patterns. For any pair~$p$ and~$p'$ of frequent path patterns found in Steps~\ref{step:simple} and~\ref{step:reachability}, we generate candidates for rules~$p \Rightarrow p'$ and~$p' \Rightarrow p$ and check whether there are~$\theta$ common sources that match both~$p$ and~$p'$; if so, we compute rule measures.
\end{enumerate}

This baseline algorithm reduces candidates by applying Lemma~\ref{lemma:antimonotonicity_1}. However, it still examines a large number of path patterns and rule candidates. To further reduce these candidates, we employ the optimization strategies discussed next.

%% file: 04.3.optim.tex
\subsection{\method}\label{sec:opt}

\pkk{To achieve efficiency in \name, we develop \method, an algorithm that lessens the explored path patterns and rules by bound-based pruning and enhanced candidate generation.}

\noindent
{\bf Bound-based pruning.} Anti-monotonicity properties eliminate candidates of path patterns whose prefixes are infrequent. Here, we introduce two upper bounds on the number of vertices matching a path pattern: (1) an upper bound on the number of vertices matching~$\langle p, \elabel, A \rangle$, and (2) an upper bound on the number of vertices matching~$\langle A, \elabel, p \rangle$, where~$p = \langle A_0, \elabel_0, A_1, \cdots, A_{n-1}, \elabel_{n}, A_n\rangle$ is an arbitrary path pattern, $\elabel$ is an edge label, and~$A$ is an attribute set. We prune results with upper bound below~$\minsup$.

To derive such upper bounds, we use (i) the set of edges with label~$\elabel$ that connect to a vertex whose attribute set includes~$A$, $\edges(A, \elabel) = \{(v, \elabel_e, v') | (v, \elabel_e, v') \in \edges \land A(v') \supseteq A \land \elabel_e = \elabel\}$; and (ii) the set of vertices whose attribute set covers~$A$ and which have an out-going edge with label~$\elabel$, $\vertices(A, \elabel) = \{v | \exists (v, \elabel_e, v') \in \edges \land A(v) \supseteq A \land \elabel_e = \elabel\}$.
The following lemmata specify our bounds:

\begin{lemma}\label{lemma:maxpath1}
\pk{Given length~$i > 0$, edge label~$\elabel$, and attribute set~$A$, the number of vertices that match a path pattern of length~$i$ ending with~$\langle \elabel, A \rangle$
is upper-bounded by~$|\edges(A, \elabel)| \cdot {d_m}^{i-1}$ where~$d_m$ is the maximum in-degree in the graph.} 
\end{lemma} 


\begin{lemma}\label{lemma:maxpath2}
\pk{Given length $i > 0$, attribute set~$A$, and edge label~$\elabel$, the number of vertices that match path pattern of the form~$p = \langle \ldots, \elabel_{i-2}, A, \elabel, A_i, \ldots \rangle$ is upper-bounded by~$|\vertices(A, \elabel)| \cdot {d_m}^{i-1}$ where~$d_m$ is the maximum in-degree in the graph.}
\end{lemma}


We use these lemmata to prune candidate path patterns.
We prune patterns of length~$i$ with suffix~$\langle \elabel, A \rangle$ if~$|\edges(A, \elabel) | \cdot {d_m}^{i-1} < \minsup$. We collect the set of single attributes that pass this pruning when combined with edge label~$\elabel$ and path length~$i$ as~$\attributes_{T_i}(\elabel) =\{ a \in \attributes \mid |\edges(a, \elabel)|\cdot {d_m}^{i-1} \geq \minsup \}$.
Likewise, we prune patterns of length~$i$ with attribute set~$A$ whose~$\elabel_{i-1}$ is~$\elabel$, if~$|\vertices(A, \elabel)| \cdot {d_m}^{i-1} < \minsup$.
We collect the set of edge labels that pass this pruning when combined with single attribute~$a \in \attributes$ and path length~$i$ as~$\elabels_{T_i}(a) =\{\elabel \in \elabels \mid |\vertices(a, \elabel)| \cdot {d_m}^{i-1} \geq \minsup  \}$. The case of $\elabel^*$ is equivalent to $\elabels_{T_1}(a)$.

Thus, bound-based pruning reduces the candidate edge labels and attribute sets for addition to frequent path patterns.




\noindent
{\bf Enhanced candidate generation.} Using the two anti-monotonicity properties and bound-based pruning, we eliminate candidate path patterns that (1)~have an infrequent prefix, (2)~are dominated by infrequent path patterns, and (3)~fall short of bound-based pruning. Pruned patterns are not included in the candidates. Here we introduce optimized candidate generation, which extends path patterns in the following ways.

\begin{itemize}
\item {\it Vertical}: starting with path patterns of length zero (i.e., frequent attributes), we extend them to length~\minlen{} following Lemma~\ref{lemma:antimonotonicity_1} and adding suffixes that are not pruned by either Lemma~\ref{lemma:maxpath1} or Lemma~\ref{lemma:maxpath2}.
\item {\it Horizontal}: starting with unit path patterns (whose attribute sets include only a single attribute each), we combine them to form frequent path patterns by applying Lemma~\ref{lemma:antimonotonicity_2}.
\end{itemize}

We revise our algorithm of Section~\ref{sec:baseline} to use enhanced candidate generation. We explain the modifications in each step. 

\begin{enumerate}
\item {\bf Frequent attribute set discovery.} We additionally find the~$\attributes_T$ and~$\elabels_T$ sets.

\item {\bf Frequent simple path pattern discovery.} We first enumerate unit path patterns of length~1, \emph{vertically} extending~$\pathpatterns_0$ by adding an edge label~$\in \elabels_T$ and an attribute~$\in \attributes_T$.
After checking the frequency of all unit path patterns of length~1, we \emph{horizontally} combine pairs of frequent path patterns to obtain new path patterns with more than one attributes. This horizontal extension drastically reduces candidates because, by Lemma~\ref{lemma:antimonotonicity_2}, if any of two paths is not frequent, the combined path patterns are not frequent either. We repeat vertical and horizontal extensions until we obtain frequent path patterns of length~\minlen{}. \label{step:simple2}

\item {\bf  Frequent reachability path pattern discovery.} We restrict the candidates for~$\elabel^*$ to edge labels in~$\elabels_T$ and those for~$A_1$ to~$\attributes_T$. We find frequent reachability path patterns whose attribute sets include a single attribute and then combine pairs of such path patterns to obtain complex reachability path patterns, following Lemma~\ref{lemma:antimonotonicity_2}. \label{step:reachability2}

\item {\bf Rule discovery.} We generate candidate rules utilizing path patterns found to be frequent in Steps~\ref{step:simple2} and~\ref{step:reachability2}, applying both Lemma~\ref{lemma:antimonotonicity_1} and Lemma~\ref{lemma:antimonotonicity_2}. We first search for frequent rules that combine unit path patterns of length~1 and then extend those patterns vertically and horizontally. For a frequent rule~$p \Rightarrow p'$, we generate candidates~$p_v \Rightarrow p'$, $p_h \Rightarrow p'$, $p \Rightarrow p'_v$, and~$p \Rightarrow p'_h$, where~$p_v$ ($p_h$) is a frequent vertical (horizontal) extension of~$p$, found in Steps~\ref{step:simple2} and \ref{step:reachability2}. 
\end{enumerate}

\method reduces the number of candidates while mining frequent path patterns and rules while guaranteeing correctness.

%% file: 04.4.auxiliary.tex
\subsection{Auxiliary data structure}\label{sec:aux}

\pk{To facilitate efficient rule discovery,
we maintain a data structure that stores, for each vertex, a list of matched path patterns and the targets of their corresponding paths. We extend paths using this data structure without searching from scratch. In addition, we maintain pairs of dominating and dominated path patterns, so as to generate rule candidate so as to avoid generating rule involving them. In effect, for a frequent rule~$p \Rightarrow p'$, we obtain~$p_v$, $p_h$, $p'_v$, and~$p'_h$ via this auxiliary data structure.}

%% file: 04.5.pseudocode.tex
\subsection{Complexity analysis}\label{sec:pseudo}

\pk{ We now analyze the time and space complexity of our algorithm. We denote the sets of candidate attribute sets as~$\candidates^A_j$, simple path patterns as~$\candidates^S_j$, reachability path patterns as~$\candidates^*_j$, and rules as~$\candidates^R_j$, where~$j$ denotes the iteration.

\noindent
\underline{\it{Time complexity}}: The frequent attribute discovery incurs the same time complexity of traditional algorithms such as Apriori algorithm~\cite{agrawal1993mining}, i.e., $O(|\attributes| + |\edges| |\vertices| + \sum_{j=2}^{|\attributes|} |\vertices| |\candidates^A_j|)$. The time complexity of frequent simple path pattern discovery depends on the number of frequent attributes and edge labels; it is~$O(|\elabels_{T}| |\attributes_{T}|^2 + \sum_{j=2}^{|\attributes|} |\vertices| |\candidates^S_j|)$. The frequent reachability path discovery step incurs a similar complexity as the previous step, while it also performs bread-first search; thus, it needs~$O(|\elabels_T|(|\vertices| + |\edges|) + |\elabels_T| |\attributes_T|^2+\sum_{j=2}^{|\attributes|}|\vertices| |\candidates^*_j|)$. The rule discovery step combines pairs of path patterns and extends the patterns in found rules. It takes a worst-case time of~$O(|\vertices| |\pathpatterns_1| |\pathpatterns^*| + \sum_{j=2}^{|\attributes|} |\vertices| |\candidates^R_j|)$. 
In total, time complexity is~$O(|\vertices||\edges|+|\elabels_T|(|\vertices| + |\edges| + |\attributes_{T}|^2)+\sum_{j=2}^{|\attributes|}|\vertices|(|\candidates^A_j| + |\candidates^S_j| + |\candidates^*_j|+|\candidates^R_j|))$,
which highly depends on the size of the candidates.

\noindent
\underline{\it{Space complexity}}: The space complexity of the algorithm is~$O(|\vertices| + |\edges| + |\attributes| + |\candidates^A| + |\candidates^S| + |\candidates^*| + |\candidates^R| + |\vertices|(\sum_{i=0}^k|\pathpatterns_i| + |\pathpatterns^*| + |\rules|))$, where~$|\candidates^A|$, $|\candidates^S|$ $|\candidates^*|$, and $|\candidates^R|$ are the maximum sizes of $|\candidates^A_j|$, $|\candidates^S_j|$ $|\candidates^*_j|$, and $|\candidates^R_j|$ over iterations~$j$, respectively. Practically, memory usage does not pose an important problem on commodity hardware.}

%% file: 05-approx.tex
\section{Approximation Techniques}\label{sec:approximation}

\method reduces the candidates of path patterns for efficient mining. However, it still does not scale to large graphs as it has to exactly compute the frequency of all unpruned path patterns, whether they are frequent or not. We present two approximation methods to reduce the computation costs for checking the frequency.

\subsection{Approximate Candidate Reduction}\label{sec:CR}

\pk{The bound-based pruning in Lemma~\ref{lemma:maxpath1} computes upper bounds using the maximum in-degree, hence may retain candidates that are not likely to be involved in frequent path patterns. Our first approximation strategy aims to eliminate such candidates in~$A_T$.}

\noindent
\pk{{\bf Method}. Given a candidate reduction factor~$\psi$ ($0 \leq \psi \leq 1$), we remove the suffix~$\langle \elabel, A \rangle$ from~$A_T$ if~$|\edges(A, \elabel)| \cdot {(d_{m})}^{\psi(i-1)} < \minsup$. Assuming a power-law edge distribution, this approximate candidate reduction effectively reduces the candidates with a small expected accuracy loss, since most vertices have much smaller in-degrees than the maximum in-degree used by Lemma~\ref{lemma:maxpath1} to prune candidates.

\noindent
{\bf Theoretical analysis}. Our approximate candidate reduction method can effectively reduce candidates, yet it may also cause false negatives, eliminating path patterns that are frequent. We discuss the probability that a frequent path pattern is eliminated.

\begin{theorem}\label{theorem:candidatereduction}
Given a frequent path pattern~$p$ of length~$i$ ending with~$\langle \elabel, A \rangle$,  the probability that~$p$ is pruned is~$P\left( \minsup > |\vertices(p)| \frac{{d_{m}}^{\psi(i-1)}}{ 
c}\right)$, 
where~$c$ is the ratio of $|\vertices(p)|$ to $|\edges(A,\elabel)|$, i.e., $|\vertices(p)| = c|\edges(A,\elabel)|$.
\end{theorem}


This theorem indicates that the probability of false negative decreases as~$|\vertices(p)|$ and the maximum in-degree grow. Therefore, we rarely miss a path matching many vertices in a large graph.}



\input{figures/runtime_vs_minsup}

\subsection{Stratified Vertex Sampling}\label{sec:sampling}

\pk{Sampling effectively reduces the computation cost in many data mining tasks~\cite{fan2022discovering, lee1998sampling, fan2022parallel}. We propose a sampling method that picks a set of vertices to work with, aiming to reduce the computation cost to find matched vertices and theoretically analyze its approximate accuracy.}

\pk{We do not reduce the graph itself, as in~\cite{fan2017big}, as then it would be hard to guarantee accuracy. Instead, as our algorithm focuses on frequent \emph{path} patterns, rather than \emph{subgraphs}, 
we can afford to reduce the number of source vertices by vertex sampling.}

\noindent
{\bf Method}. \pk{We need a sampling strategy that preserves accuracy as much as possible. 
To achieve that, we use {\it stratified} sampling according to attributes of vertices~\cite{thompson2002sampling}. We group vertices into \emph{strata} according to their attribute sets and remove vertices that have no frequent attributes because they do not contribute to frequent path patterns. From each stratum, we pick vertices with sampling ratio~$\rho$. We estimate the frequency of~$p$ as follows:
\begin{equation}
     \widetilde{|\vertices(p)|} = \frac{|\vertices_s(p)|}{\rho}.
\end{equation}
$\vertices_s$ denotes the set of sampled vertices in strata related to~$A_0$ of~$p$.

\noindent
{\bf Theoretical analysis}. The accuracy of sampling is expressed by variance. The variance of our sampling strategy is:
\begin{equation}
    s^2=\frac{\sum_{v_i\in \vertices_s} (x_i -\overline{x})}{\rho|\vertices_s|-1}.
\end{equation}
where~$x_i = 1$ if~$v_i$ matches~$p$, otherwise~0. $\overline{x}$ is $\frac{|\vertices_s(p)|}{ |\vertices_s| }$.

The confidence interval is then the following:
\begin{equation}
     |\vertices_s|(\overline{x} - z \cdot \frac{s}{\sqrt{\rho|\vertices_s|}}) < |\vertices(p)| < |\vertices_s|(\overline{x} + z \cdot \frac{s}{\sqrt{\rho|\vertices_s|}})
\end{equation}
where~$z$ indicates the $z$-value for a confidence interval.

Contrary to candidate reduction, vertex sampling may cause false positives, that is, infrequent path patterns could be reported as frequent path patterns. Yet it can reduce the computation costs, even in cases where the candidate reduction does not.}


%% file: figures/runtime_vs_minsup.tex
\begin{figure*}[!t]
  \subfloat[Nell]{\epsfig{file=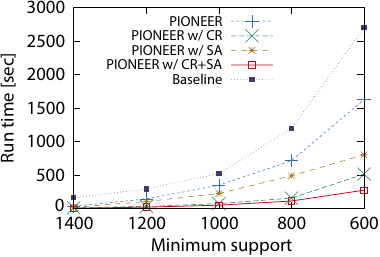,width=0.25\linewidth}
  }
    \subfloat[MiCo]{\epsfig{file=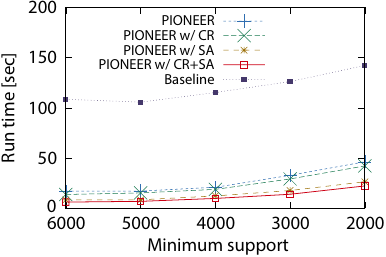,width=0.25\linewidth}
  }
  \subfloat[DBpedia]{\epsfig{file=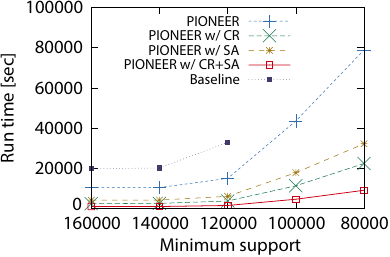,width=0.25\linewidth}
  }
  \subfloat[Pokec]{\epsfig{file=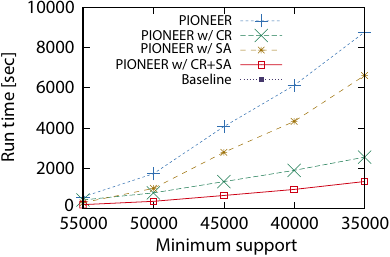,width=0.25\linewidth}
  }
\vspace{-5mm}
\caption{Impact of minimum support~$\minsup$ on run time; missing points indicate that a method did not finish within~24 hours; on Pokec, the Baseline did not finish within~24 hours for all minimum support values.}
\label{fig:runtime_minsupport}
\vspace{-3mm}
\end{figure*}

%% file: 06-parallel.tex
\section{Parallelization}\label{sec:parallelization}

\pk{We accelerate \method by parallelization. Given a number of threads~$N$, we partition the set of vertices into~$N$ subsets to balance the computing cost among threads in terms of the frequency of vertex attributes and adjacent edges. We describe how we estimate computing costs in the following.}

\noindent
{\bf Cost estimation.} \pk{The cost to find matched vertices increases with the number of matched paths from each vertex. First, we prune vertices that do not have target attributes or outgoing edges with target edge labels, as these vertices have no matched frequent path patterns, as the following lemma specifies.}

\begin{lemma}
\pk{If~$v \in \vertices$ has no target attributes and no outgoing edges with target labels, then~$v$ has no matched path patterns.}
\end{lemma}

\pk{Among non-pruned vertices, the more frequent their attributes and outgoing edges are, the more matched path patterns they may match. We estimate the cost~$C(v)$ of a vertex~$v$ as:
\begin{equation}
C(v) = d_T(v) \cdot |A_T(v)|
\end{equation}
where~$d_T(v)$ is the number of out-going edges with target edge labels and~$|A_T(v)|$ is the number of target attributes on~$v$. This estimation is~$O(1)$ considering the numbers of edges and attributes on~$v$ as constants.}

\noindent
{\bf Partitioning.} \pk{We partition the set of vertices into~$N$ subsets according to estimated costs by a greedy algorithm. We sort vertices in ascending order of their costs and repeatedly assign the unassigned vertex of the largest cost to the thread with the smallest sum of assigned vertex costs; we do not assign a vertex to a thread that is already assigned~$\frac{|\vertices|}{N}$ vertices. The time complexity of this algorithm is~$O(N |\vertices|\log |\vertices|)$.}


%% file: 07-experiment.tex
\section{Experimental Study}\label{sec:experiment}

We conduct an experimental study on \method to assess (1) its efficiency, (2) its scalability, and (3) the accuracy of our approximations. We also assess (4) the effectiveness of \name. We implemented all algorithms in C++ and ran experiments on a Linux server with~512GB of main memory and an Intel(R) Xeon(R) CPU E5-2699v3 processor.
Some experimental settings and results (e.g., length $k$ and approximation factors) are in the appendix.

\subsection{Experimental Setting}



\noindent
{\bf Dataset}. We use four real-world graphs with edge labels and vertex attributes: 
 knowledge graph \textsf{Nell}~\cite{carlson2010toward},
coauthorship information network \textsf{MiCo}~\cite{auer2007dbpedia},
knowledge graph extracted from Wikipedia \textsf{DBpedia}~\cite{auer2007dbpedia}, and
social network service \textsf{Pokec}.
We also use two types of synthetic graphs, \textsf{uniform} and \textsf{exponential}. They differ on how we generate edges; in \textsf{uniform}, we generate edges between randomly selected vertices following a uniform distribution, while in \textsf{exponential} we follow an exponential distribution. We vary the number of vertices from~1M to~4M and generate a fivefold number of edges (i.e, 5M to~20M). 
Table~\ref{table:datasets} presents statistics on the data.

\input{tables/statistics}

\noindent
{\bf Compared methods}. \pkk{We assess a baseline and four variants of \method.} \textsf{Baseline} is the algorithm of Section~\ref{sec:baseline} without any optimization strategy. \textsf{\method} is our algorithm using the strategies of Section~\ref{sec:opt}. \textsf{\method w/ CR} using the candidate reduction of Section~\ref{sec:CR}, while \textsf{\method w/ SA} uses the stratified sampling of Section~\ref{sec:sampling}, and \textsf{\method w/ CR+SA} employs both. All algorithms are parallelized by the technique of Section~\ref{sec:parallelization}. Further, we compare the run time of CSM-A~\cite{prateek2020mining}\footnote{https://github.com/arneish/CSM-public}, a method that approximately finds the top-$k$ frequent correlated subgraph pairs, to \method, even though the output of CSM-A is different from that of PARM. 
Notably, other extant graph association rule mining methods~\cite{fan2015association, fan2017big, wang2020extending, fan2022discovering} do not address the \name problem and do not have open codes.


\noindent
{\bf Parameters}. 
We set minimum support threshold~1\,000 on Nell, 4\,000 on MiCo, 120\,000 on DBpedia, and 45\,000 on Pokec, respectively; both the candidate reduction factor~$\psi$ and the sampling rate~$\rho$ are set to~0.4; we set the maximum path length~$k=2$, and use 32~threads. We compute absolute support, relative support, confidence, and lift.
We vary these parameters to evaluate their impacts while using the above values as default parameters.



\subsection{Efficiency}


\noindent
{\bf Varying minimum support~$\minsup$}. 
Figure~\ref{fig:runtime_minsupport} plots run time vs. minimum support on each data. 
The minimum support directly affects the number of rules to be discovered, hence computational cost. 
As the minimum support falls, the number of candidate path patterns, and hence run time, grows. Our algorithms outperform the baseline in terms of efficiency. In Pokec, the baseline did not finish within~24 hours due to the large number of attributes.

We observe that the enhanced candidate generation, employed in \textsf{Ours}, is effective in reducing candidate path patterns. Further, our approximation methods reducing the number of candidate path patterns and processed vertices by sampling further reduce the computational cost. Our algorithm employing both of these approximation methods consistently achieves the lowest runtime. Regarding the comparison between those two methods, our algorithm with candidate reduction is more effective than that with sampling on Nell, DBpedia, and Pokec, as the number of candidates is often larger than the number of vertices on those data. In MiCo, on the other hand, candidate reduction is not so effective because it does not reduce the candidates with the default~$\psi=0.4$.




\noindent
{\bf Comparison to CSM-A}. We juxtapose the run time of \method to that of CSM-A. 
Figure~\ref{fig:comparisoncsm} presents our results. On Pokec, \method is much faster than CSM-A, indicating that CSM-A is less scalable in graph size. On DBpedia \method is less efficient, as DBpedia has a large average number of attributes per vertex, yielding a larger search space for \method than for CSM-A (i.e., we reduce the number of attributes per vertex to one for CSM-A).

\input{figures/runtime_csm}

\noindent
{\bf Varying path length~$k$}. As the path length~$k$ grows, the candidate path patterns increase, thus computation cost grows.
On Nell, the number of rules drastically increases as~$k$ grows, hence candidate reduction becomes more effective than sampling when~$k=3$. With~$k=4$, due to the growing number of rules, our algorithms did not finish within~24 hours. Arguably, to effectively find path association rules, we need to either increase the minimum support or reduce approximation factors. On DBpedia the number of path patterns does not grow when~$k>2$, so run time does not increase either, as some vertices have no outgoing edges, thus most path patterns have length~2. In Pokec, when the $k>2$, the number of reachability patterns increases, thus the run time increases. Overall, our approximation methods work well when~$k$ is large, as the run time gap between the exact and approximate algorithms widens.

\smallskip\noindent
{\bf Varying approximation factors}. The approximation factors~$\psi$ for candidate reduction and~$\rho$ for sampling indicate the degree of approximation. As both decrease, the extent of approximation increases. 
Notably, when approximation factors are small, run time is low, with the partial exception of Nell. On Nell, the run time of \textsf{\method w/ SA} is high when~$\rho = 0.2$ due to many false positives that burden the rule discovery step, as Nell is a small graph compared with DBpedia and Pokec. On the other hand, candidate reduction does not increase the number of found frequent path patterns and rules, so the run time of \textsf{\method w/ CR} grows with the approximation factor. In conclusion, sampling effectively reduces run time in large graphs, yet it may not be effective on small graphs.

\subsection{Scalability}

\noindent
{\bf Varying the number of threads}. Table~\ref{table:numberthreads} shows runtime vs. the number of threads; the runtime of our algorithms decreases almost linearly as the number of threads rises, especially on larger data.

\input{tables/runtime_vs_threads}

\noindent
{\bf Memory usage}. Table~\ref{table:memoryusage} presents the memory usage of our algorithms. DBpedia raises the highest memory requirements, Pokec the lowest, indicating that memory use depends on the number of candidates more than on graph size. Our algorithms reduce memory usage by reducing candidates, while vertex sampling further reduces memory use by reducing path patterns to be stored for each vertex. When the number of frequent path patterns is large, the approximation methods effectively reduce memory use, as on Nell, confirming our space complexity analysis.

\input{tables/memory}

\noindent
{\bf Graph size}. Figure~\ref{fig:runtime_graphsize} depicts run time on uniform and exponential synthetic graphs with~0.01 as the relative minimum support. The results suggest that run time grows linearly. Thus, our algorithms are highly scalable to large graphs. On uniform data, vertex sampling is more effective than candidate reduction, while on exponential data, candidate reduction proves to be more effective, since it works best when the maximum degree deviates far from the average.

\input{figures/graph_size}

\subsection{Accuracy}

We evaluate our approximation methods vs. exact ones in terms of recall, the fraction of true frequent rules that are found, and precision, the fraction of found frequent rules that are true. Table~\ref{table:accuracy_factor} shows the results. Both recall and precision are quite high in MiCo, DBpedia, and Pokec. In DBpedia, they do not fall even if we set~$\psi$ and~$\rho$ to~0.2. In Nell, such measures fall as the approximation factors decrease. When applying candidate reduction, the number of matched vertices and the maximum in-degrees are small in Nell, so the missing probabilities become large by Lemma~\ref{theorem:candidatereduction}. Still, as the difference between the maximum and average in-degrees is large in DBpedia and Pokec, we may reduce candidates without compromising accuracy. Vertex sampling leads to more inaccuracies on Nell and MiCo, which are small compared to DBpedia and Pokec; our approximation methods work well on large graphs.

\input{tables/approximation_factors}

\subsection{Effectiveness}

As no previous work finds path association rules and \name does not target quantitatively measurable outcomes, we evaluate its effectiveness on \emph{bias checking} and \emph{knowledge extraction}.

\noindent
{\bf Bias checking}. Detecting data bias is essential for building machine learning models that avoid algorithmic bias. In particular, gender biases in datasets are evaluated in several works~\cite{levy-etal-2021-collecting-large,imtiaz2019investigating}
We apply \name to check gender bias in Pokec. We evaluate biases among female and male persons with respect to the education of their friends, comparing the support and confidence of rules of the form~$\langle \{\labelfont{gender} \} \rangle \Rightarrow \langle \{ \}, \labelfont{follows}, \{\labelfont{education:A} \} \rangle$. We found~671 rules of the above pattern with~$\minsup=500$. The sums of absolute supports for men and women are~296\,855 and~466\,725, respectively, and the sums of confidences are~2.45 and~1.43, respectively. Since the numbers of vertices with men and women are~804\,327 and~828\,275, i.e., there are more women than men, this result suggests that men are more likely to have friends who registered their educations than women. Besides, the sum of absolute supports for men who unset their ages is three times larger than that for women who unset their ages (151\,240 vs 53\,370). It follows that Pokec contains a gender bias and one should take care when using it to train ML models.

\noindent {\bf Knowledge extraction}. 
\pkk{We discuss experimentally found rules.}

\noindent
\underline{Nell}. \pkk{We discovered a rule indicating that if a `television station' (vertex~$v$) is part of a `company' employing a `CEO', then~$v$ is also part of a `company' employing a `professor'. Notably, in Nell, there are no vertices associated with both a CEO and a professor, and companies employing both roles are rare. This rule thus elucidates the operations of large organizations.}

\noindent
\underline{DBpedia} We identified the rule that individuals with a registered birthplace are usually born in \emph{populated} areas, with confidence of~0.997, which implies that births in unpopulated places exist. We found three such types: (a)~significant buildings like palaces; (b)~areas, such as lakes, that are now considered unpopulated; and (c)~incomplete data, especially in Africa and South America. These findings highlight areas for DBpedia's expansion. We found~60 places that miss a label of ``populatedPlace'' among 302 places included in the rules, e.g., Misiones Province, Al Bahah Province, and Oaxaca. The rest 242 places were significant buildings and areas. PARM helps to explainably complement missing labels.

\noindent
\underline{Pokec}. \pkk{We mined the rule that female users who have \emph{not} set their ages and are within~\emph{2 hops} of other female users are \emph{likely} to follow male users who also have \emph{not} set their ages, with a confidence of~0.63. However, the confidence of this rule is \emph{lower} compared to some other rules that have confidence over~0.8. This pattern sheds light on social dynamics in the network. We also observed a rule involving reachability; 
namely, \emph{men in their~20s} who are within~4 hops of men who \emph{unset} their ages are also reachable within~4 hops from \emph{women in their~20s}, with a confidence of~0.999. This rule illustrates how \name facilitates connectivity analysis in social networks.}


%% file: tables/statistics.tex
\begin{table}[!h]
\vspace{-2mm}
\caption{Data statistics.}\label{table:datasets}
\vspace{-3mm}
\scalebox{1}{
\begin{tabular}{lrrrrr}
\hline
          &\multicolumn{1}{c}{$|\vertices|$} & \multicolumn{1}{c}{$|\edges|$} & \multicolumn{1}{c}{$|\elabels|$} & \multicolumn{1}{c}{$|\attributes|$} & \multicolumn{1}{c}{Avg. Attr.}\\ \hline
Nell      & 46,682      &   231,634     & 821       & 266       & 1.5       \\
MiCo   & 100,000   &   1,080,298   & 106       & 29       & 1.0       \\ 
DBpedia   & 1,477,796   &   2,920,168   & 504       & 239       & 2.7       \\ 
Pokec     & 1,666,426   &  34,817,514 & 9            & 36,302   & 1.1      \\
Synthetic  & 1M -- 4M   & 5M -- 20M & 500      &  500   & 2.0      \\ 
\hline
\end{tabular}}
\vspace{-2mm}
\end{table}

%% file: figures/runtime_csm.tex
\begin{figure}[!t]
\vspace{-1mm}
    \centering
    \includegraphics[width=0.8\linewidth]{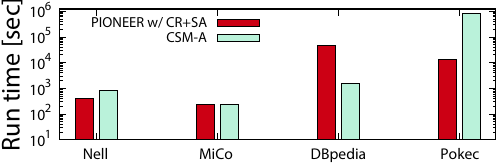}
\vspace{-4mm}    
\caption{Run time comparison with CSM-A}\label{fig:comparisoncsm}
\vspace{-4mm}
\end{figure}

%% file: tables/runtime_vs_threads.tex
\begin{table}[!t]
\vspace{-2mm}
\caption{Impact of number of threads on run time [sec]}\label{table:numberthreads}
\vspace{-4mm}
\scalebox{1.0}{
\begin{tabular}{ll|rrr}
\hline
\multirow{2}{*}{Dataset}& \multirow{2}{*}{Method}& \multicolumn{3}{|c}{Number of threads}\\
&    & \multicolumn{1}{c}{8}     & \multicolumn{1}{c}{16} & \multicolumn{1}{c}{32} \\ \hline
\multirow{2}{*}{Nell}&\method    &     754.6 &    496.2 &     348.7 \\
&\method w/ CR+SA                &      74.0 &     63.2 &      51.7 \\\hline
\multirow{2}{*}{MiCo}&\method    &      70.1 &     37.5 &      21.0 \\
&\method w/ CR+SA                &      31.9 &     17.4 &      10.0 \\\hline
\multirow{2}{*}{DBpedia}&\method & 59\,603.0 & 29\,971.0 & 15\,028.0 \\
&\method w/ CR+SA                &  6\,037.0 &  3\,042.0 &  1\,529.0 \\\hline
\multirow{2}{*}{Pokec}&\method   & 10\,039.0 &  6\,197.0 &  4\,038.0\\
&\method w/ CR+SA                &  1\,981.0 &  1\,103.0 &     634.0\\
\hline
\end{tabular}
}
\vspace{-2mm}
\end{table}

%% file: tables/memory.tex
\begin{table}[!t]
\vspace{-2mm}
\caption{Memory usage [GB]}\label{table:memoryusage}
\vspace{-4mm}
\begin{tabular}{l|rrrr}
\hline
& \multicolumn{1}{c}{Nell} & \multicolumn{1}{c}{MiCo}& \multicolumn{1}{c}{DBpedia} & \multicolumn{1}{c}{Pokec} \\ \hline
\method & 11.1 & 0.57& 23.9 & 6.4\\
\method w/ CR & 4.2& 0.55 &21.5& 6.4 \\
\method w/ SA &5.6& 0.56 & 23.8 &  6.1\\
\method w/ CR+SA &2.4& 0.53 &21.5 & 6.1 \\
\hline
\end{tabular}
\vspace{-2mm}
\end{table}

%% file: figures/graph_size.tex
\begin{figure}[!t]
  \subfloat[Uniform]{\epsfig{file=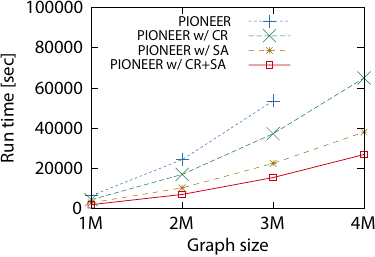,width=0.5\linewidth}
  }
\subfloat[Exponential]{\epsfig{file=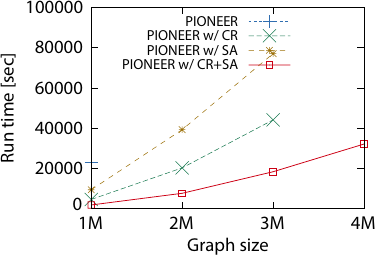,width=0.5\linewidth}
  }
\vspace{-4mm}
\caption{Scalability to graph size.}\label{fig:runtime_graphsize}
\vspace{-4mm}
\end{figure}

%% file: tables/approximation_factors.tex
\begin{table}[!t]
\vspace{-2mm}
\caption{Impact of approximation factors on accuracy.}\label{table:accuracy_factor}
\vspace{-2mm}
\begin{subtable}[h]{1.0\linewidth}
\scalebox{1}{
\begin{tabular}{l|ll|ll|ll}
\hline
\multirow{2}{*}{$\psi$ and $\rho$}& \multicolumn{2}{|c|}{CR}& \multicolumn{2}{|c|}{SA} & \multicolumn{2}{|c}{CR+SA}\\ 
             & Recall     & Precis & Recall& Precis     & Recall & Precis\\ \hline
0.2	&0.046	&1.0	&0.9999	&0.495	&0.046	&0.500\\
0.4	&0.233	&1.0	&0.9999	&0.999	&0.233	&0.998\\
0.6	&0.532	&1.0	&0.9999	&0.9999	&0.532	&0.9999\\
0.8	&0.797	&1.0	&0.9999	&0.9999	&0.797	&0.9999\\
\hline
\end{tabular}
}
\caption{Nell}\label{table:accuracy_factor_nell}
\end{subtable}
\begin{subtable}[h]{1.0\linewidth}
\scalebox{1}{
\begin{tabular}{l|ll|ll|ll}
\hline
\multirow{2}{*}{$\psi$ and $\rho$} & \multicolumn{2}{|c|}{CR}& \multicolumn{2}{|c|}{SA} & \multicolumn{2}{|c}{CR+SA}\\ 
    & Recall     & Precis & Recall& Precis     & Recall & Precis\\ \hline
0.2	&1.0	&1.0	&1.0	&0.978	&1.0	&0.978\\
0.4	&1.0	&1.0	&1.0	&0.882	&1.0	&0.882\\
0.6	&1.0	&1.0	&1.0	&0.938	&1.0	&0.938\\
0.8	&1.0	&1.0	&0.978	&1.0	&0.978	&1.0\\
\hline
\end{tabular}
}
\caption{MiCo}\label{table:accuracy_factor_mico}
\end{subtable}
\begin{subtable}[h]{1.0\linewidth}
\scalebox{1}{
\begin{tabular}{l|ll|ll|ll}
\hline
\multirow{2}{*}{$\psi$ and $\rho$} & \multicolumn{2}{|c|}{CR}& \multicolumn{2}{|c|}{SA} & \multicolumn{2}{|c}{CR+SA}\\ 
             & Recall     & Precis & Recall& Precis     & Recall & Precis\\ \hline
0.2	&1.0	&1.0		&1.0	&1.0	&1.0	&1.0\\
0.4	&1.0	&1.0		&1.0	&1.0	&1.0	&1.0\\
0.6	&1.0	&1.0		&1.0	&1.0	&1.0	&1.0\\
0.8	&1.0	&1.0		&1.0	&1.0	&1.0	&1.0\\
\hline
\end{tabular}
}
\caption{DBpedia}\label{table:accuracy_factor_dbpedia}
\end{subtable}
\begin{subtable}[h]{1.0\linewidth}
\scalebox{1}{
\begin{tabular}{l|ll|ll|ll}
\hline
\multirow{2}{*}{$\psi$ and $\rho$} & \multicolumn{2}{|c|}{CR}& \multicolumn{2}{|c|}{SA} & \multicolumn{2}{|c}{CR+SA}\\ 
             & Recall     & Precis & Recall& Precis     & Recall & Precis\\ \hline
0.2	&1.0	&1.0	&0.934	&0.983	&0.934	&0.983\\
0.4	&1.0	&1.0	&0.967	&0.983	&0.967	&0.983\\
0.6	&1.0	&1.0	&0.984	&0.986	&0.984	&0.984\\
0.8	&1.0	&1.0	&0.984	&0.984	&0.984	&0.984\\
\hline
\end{tabular}
}
\caption{Pokec}\label{table:accuracy_factor_pokec}
\end{subtable}
\vspace{-5mm}
\end{table}

%% file: 08-conclusion.tex
\section{Conclusion}\label{sec:conclusion}

\pkk{We introduced the problem of \emph{path association rule mining} (\name), which finds regularities among path patterns in a single large graph, and developed \method, an algorithm for \name. Our experimental study confirms that \method efficiently finds interesting patterns. In the future, we aim to extend our method to find top-$k$ rules by novel measures capturing features of path patterns, authenticate top-$k$ results as in~\cite{papadopoulos11}, extend patterns to subgraphs, and add a step that computes other measures; we also aim to examine how the mining of path association rules can inform the discovery of outstanding facts in knowledge graphs~\cite{yang21} and how the mining of rules can be informed by fairness measures, such as those of~\cite{giannakopoulos15, tziavelis19, tziavelis20, wu23}.}

%% file: 90-appendix.tex
\appendix

\section{Notations}
Table~\ref{table:notation} gathers the notations we employ.

\input{tables/notations}

\section{Related work (continued)} 

{\bf Path queries}.
Since \name aims to discover regularities between paths, it relates to path queries on graphs, an active field of study~\cite{AnglesABHRV17, AnglesRV19, yuya2022, valstar2017reachability}.
However, path queries do not aim to find \emph{frequent} path patterns.
In addition, SPARQL systems can find the number of paths that match given patterns, but it does not focus on frequent mining and measures of data mining such as confidence and lifts, so they do not directely support association rule mining.

Recent real-life query logs for Wikidata~\cite{vrandevcic2014wikidata} showed that more than 90\% of queries are path patterns~\cite{bonifati2020analytical,BonifatiMT19}. Due to this fact, path patterns are often used in graph analysis instead of subgraph patterns to understand real-world relationships. In that sense, \name can be useful in real-world analysis.

\section{Proofs}

\setcounter{theorem}{0}
\setcounter{lemma}{0}

\begin{theorem}
\pk{\name is NP-hard.}
\end{theorem}
\noindent
{\it Proof sketch}: \pk{We reduce the NP-hard frequent itemset mining problem~\cite{yang2004complexity,bessiere2020computational} to \name. Given an instance of frequent itemset mining, we build, in polynomial time, a graph without edges, where each node corresponds to a transaction and a set of attributes to a set of items. Solving \name on this graph amounts to solving the frequent itemset mining problem. Thus, if \name were solvable in polynomial time, we would solve frequent itemset mining problem in polynomial time too. Hence \name is NP-hard.} \hfill{} $\square$

\begin{theorem}\label{theo:dominate}
If $p \subset p'$, $\vertices(p) \supseteq \vertices(p')$.
\end{theorem}

\begin{proof}
\pk{If~$p'$ dominates~$p$, then~$p$ is no longer than~$p'$, the edge labels~$\elabel_0, \ldots, \elabel_{n-1}$ on~$p$ and~$\elabel'_0, \ldots, \elabel'_{n-1}$ on~$p'$ are the same in the same order, and the attribute sets in~$p$ are subsets of those in~$p'$, $A_i \subseteq A'_i$ for~$i = 0 \ldots n$. Thus, any vertex that matches~$p'$ also matches~$p$, i.e., $\vertices(p) \supseteq \vertices(p')$.}
\end{proof}

\begin{lemma}\label{lemma:antimonotonicity_1}
\pk{If the prefix of a path pattern is infrequent, the whole pattern is infrequent.}
\end{lemma}

\begin{proof}
\pk{If vertex~$v$ matches a whole path pattern~$p = \langle A_0,$ $\elabel_1, \ldots,$ $\elabel_{m}, A_m, \ldots, \elabel_{n}, A_n \rangle$, then~$v$ also matches the prefix~$\langle A_0,$ $\elabel_1, \ldots,$ $\elabel_{m},$ $A_m \rangle$ where~$m < n$. In reverse, if~$v$ does not match the prefix~$\langle A_0, \elabel_1, \ldots,$ $\elabel_{m}, A_m \rangle$, then it does not match the whole pattern~$\langle A_0, \elabel_1,$ $\ldots,$ $\elabel_{m},$ $A_m, \ldots,$ $\elabel_{n}, A_n\rangle$ either. Therefore, if the prefix of a path pattern is infrequent, then the whole path pattern is infrequent.}
\end{proof}

\begin{lemma}\label{lemma:antimonotonicity_2}
\pk{If path pattern~$p$ is infrequent and~$p'$ comprises~$A_i' \supseteq A_i$ and~$\elabel_i' = \elabel_i$ for all~$i$, then~$p'$ is also infrequent.}
\end{lemma}

\begin{proof}
\pk{If vertex~$v$ does not match path pattern~$p$, then~$v$ does not match~$p'$ either, as each vertex's set of attributes in~$p'$ is a superset of the respective set in~$p$. Therefore, if~$p$ is infrequent, $p'$ is also infrequent.}
\end{proof}

\begin{lemma}\label{lemma:maxpath1}
\pk{Given length~$i > 0$, edge label~$\elabel$, and attribute set~$A$, the number of vertices that match a path pattern of length~$i$ ending with~$\langle \elabel, A \rangle$
is upper-bounded by~$|\edges(A, \elabel)| \cdot {d_m}^{i-1}$ where~$d_m$ is the maximum in-degree in the graph.} 
\end{lemma} 

\begin{proof}
\pk{$|\edges(A, \elabel)|$ is the number of edges with label~$\elabel$ that connect to a vertex whose attribute set includes~$A$. Given path pattern~$p = \langle A_0, \elabel, A \rangle$, where~$A_0$ is an arbitrary attribute set, the maximum number of vertices that may match~$p$ is~$|\edges(A, \elabel)|$. This number rises by at most a factor of~$d_m$ per unit of path length added to its prefix, hence the number of vertices matching a path pattern of length~$i$ ending with~$\langle \elabel, A \rangle$ is upper-bounded by~$|\edges(A, \elabel) | \cdot {d_m}^{i-1}$.}
\end{proof}

\begin{lemma}\label{lemma:maxpath2}
\pk{Given length $i > 0$, attribute set~$A$, and edge label~$\elabel$, the number of vertices that match path pattern of the form~$p = \langle \ldots, \elabel_{i-2}, A, \elabel, A_i, \ldots \rangle$ is upper-bounded by~$|\vertices(A, \elabel)| \cdot {d_m}^{i-1}$ where~$d_m$ is the maximum in-degree in the graph.}.
\end{lemma}

\begin{proof}
\pk{$|\vertices(A, \elabel)|$ is the number of vertices whose attribute set covers~$A$ and which have an out-going edge with label~$\elabel$. Given path pattern~$p' = \langle A, \elabel, A_1 \rangle$, where~$A_0$ is an arbitrary attribute set, the maximum number of vertices that may match~$p'$ is~$|\vertices(A, \elabel)|$. If $p'$ is extended on the suffix to~$p'' = \langle A, \elabel, A_1, \ldots \rangle$, the number of vertices that match~$p''$ does not increase by Theorem~\ref{theo:dominate}. Yet matching vertices grow by at most a factor of~$d_m$ per unit of length added to the path's prefix, as in Lemma~\ref{lemma:maxpath1}.
Hence, the number of vertices matching path pattern~$p = \langle \ldots, \elabel_{i-2}, A, \elabel, A_i, \ldots \rangle$ is upper-bounded by~$|\vertices(A, \elabel) | \cdot {d_m}^{i-1}$.}
\end{proof}

\begin{theorem}\label{theorem:candidatereduction}
Given a frequent path pattern~$p$ of length~$i$ ending with~$\langle \elabel, A \rangle$,  the probability that~$p$ is pruned is~$P\left( \minsup > |\vertices(p)| \frac{{d_{m}}^{\psi(i-1)}}{ 
c}\right)$, 
where~$c$ is the ratio of $|\vertices(p)|$ to $|\edges(A,\elabel)|$, i.e., $|\vertices(p)| = c|\edges(A,\elabel)|$.
\end{theorem}

\begin{proof}
The approximate candidate reduction removes a candidate if~$|\edges(A, \elabel)| {d_m}^{\psi (i - 1)} < \minsup$. Since~$|\vertices(p)| = c|\edges(A,\elabel)|$, it is:
\begin{eqnarray}
 & P\left( \minsup > |\edges(A,\elabel)|{{d_{m}}^{\psi(i-1)}}\right)  \nonumber\\
=& P\left( \minsup > |\vertices(p)|\frac{{d_{m}}^{\psi(i-1)}}{c}\right) \nonumber
\end{eqnarray}
\end{proof}

\begin{lemma}
\pk{If~$v \in \vertices$ has no target attributes and no outgoing edges with target labels, then~$v$ has no matched path patterns.}
\end{lemma}
\begin{proof}
\pk{If~$v$ does not match any attributes, it does not become a source of path pattern; similarly, if~$v$ has no outgoing-edges in target edges, it has no~$\elabel_1$ in path patterns; the lemma follows.}
\end{proof}

\section{Pseudo-code}
\input{algorithms/discovery}
\pk{Algorithm~\ref{alg:discovery} presents an algorithm of \method in pseudo-code. Each function named {\sf Discover*} searches for vertices matched with candidates and finds frequent patterns (Lines~5, 11, 16, and~20). Lines~2--7 comprise the step of the frequent attribute set discovery, which follows the same logic as prior work (i.e., the Apriori algorithm~\cite{agrawal1993mining}) and also computes edge labels and attributes of targets (Line~7).
Lines~8--12 make up the frequent simple path pattern discovery step, which extends already found frequent path patterns vertically and horizontally to form new candidate path patterns. Lines~13--17 show the frequent reachability path pattern discovery step, which conducts a breadth-first search for each frequent edge label from each vertex to obtains a set of reached vertices~$\vertices_T$, and then discovers a set of frequent reachability path patterns.
Eventually, the rule discovery step in Lines~18--22 first generates candidate rules that are pairs of either simple or reachability unit path patterns of length~1; it mines rules among those candidates and proceeds iteratively by extending the path patterns in the rules it discovers finds both horizontally and vertically. Lastly, it computes measures for all discovered rules.}

\section{Additional experiments}\label{app:experiment}

\pk{We use four real-world graphs with edge labels and vertex attributes:

\begin{itemize}
\item \textbf{Nell}\footnote{\url{https://github.com/GemsLab/KGist/blob/master/data/nell.zip}}~\cite{carlson2010toward} is a knowledge graph about organizations crawled from the web; it features vertex attributes such as \textsf{CEO}, \textsf{musician}, \textsf{company}, and \textsf{university}, and edge labels such as \textsf{companyceo}, \textsf{competeswith}, and \textsf{worksat}.
\item \textbf{MiCo}\footnote{\url{https://github.com/idea-iitd/correlated-subgraphs-mining}}~\cite{auer2007dbpedia} is a coauthorship information network from Microsoft; its attributes and edge labels are embedded in integers.
\item \textbf{DBpedia}\footnote{\url{https://github.com/GemsLab/KGist/blob/master/data/dbpedia.zip}}~\cite{auer2007dbpedia} is a knowledge graph extracted from Wikipedia, featuring vertex attributes such as \textsf{actor}, \textsf{award}, \textsf{person}, and \textsf{place}, and edge labels such as \textsf{child}, \textsf{spouse}, and \textsf{deathPlace}.
The vertex attribute and edge labels are types of vertices and relationships.
\item \textbf{Pokec}\footnote{\url{https://snap.stanford.edu/data/soc-pokec.html}} is a social network service popular in Slovenia; it features vertex attributes such as \textsf{age}, \textsf{gender}, \textsf{city}, and \textsf{music}, and edge labels such as \textsf{follows}, \textsf{likes}, and \textsf{locatedIn}; we divide edges labeled with \textsf{follows} into seven groups according to out-degree value: \textsf{single},  \textsf{verysmall},  \textsf{small},  \textsf{average},  \textsf{large},  \textsf{verylarge}, and \textsf{hub} are corresponding to 1, 2--4, 5--14, 15--29, 30--49, 50--100, and over 100 out-degrees, respectively.
\end{itemize}
These datasets are open publicly, so please see their Github in detail.
We also use two types of synthetic graphs, \textsf{uniform} and \textsf{exponential}. They differ on how we generate edges; in \textsf{uniform}, we generate edges between randomly selected vertices following a uniform distribution, while in \textsf{exponential} we follow an exponential distribution. We vary the number of vertices from~1M to~4M, while generating a fivefold number of edges (i.e, 5M to~20M). Table~\ref{table:datasets} presents statistics on the data.}

We here note that there are no suitable benchmarks for our problem. 
For example, a benchmark LDBC \cite{erling2015ldbc} uses synthetic graphs generated by Graphalytics~\cite{iosup2020ldbc}, but they lack vertex attributes and labeled edges.
Therefore, we use real-world graphs in various domains and simple synthetic graphs for evaluating scalability.


\input{tables/statistics}
\subsection{Efficiency in detail}\label{app:efficiency}

\noindent
{\bf Run time analysis}.
Our results indicate that the data size is not a dominant factor in the run time. For instance, DBpedia takes a longer time in Pokec even though Pokec is larger than DBpedia and the minimum support in Pokec is set smaller than that in DBpedia. To further investigate this matter, we plot, in Figure~\ref{fig:runtime_ratio}, the distribution of run time components on each step. Table~\ref{table:numberpatterns} shows the numbers of frequent attribute sets, patterns, and rules on each data set. Table~\ref{table:numberpathspatternsinrules} shows the numbers of path pattern types in rules. 

We observe that the run time distribution differs across datasets. On Nell, most time is devoted to finding rules and length-2 simple path patterns. On DBpedia, the run time is predominantly spent in finding length-2 simple patterns. On Pokec, time goes to finding all kinds of path patterns rather than rules. These distributions of run time are generally consistent with the numbers of patterns and rules in Table~\ref{table:numberpatterns}; overall, our results corroborate our time complexity analysis, where the numbers of candidates highly affect run time.

\input{figures/runtime_ratio}
\input{figures/runtime_vs_length}
\input{figures/runtime_vs_approxfactors}

\noindent
{\bf Varying path length~$k$}. As the path length~$k$ grows, the candidate path patterns increase, thus computation cost grows. Figure~\ref{fig:runtime_length} plots run time vs.~$k$, setting minimum support~$\minsup$ to~1\,800 in Nell, 140\,000 in DBpedia, and 50\,000 in Pokec. We do not show the baseline's run time as it did not finish within~24 hours for~$k \geq 3$ on all data. Notably, the run time grows vs.~$k$, with the partial exception of DBpedia. On Nell, the number of rules drastically increases as~$k$ grows, hence candidate reduction becomes more effective than sampling when~$k=3$. With~$k=4$, due to the growing number of rules, our algorithms did not finish within~24 hours. Arguably, to effectively find path association rules, we need to either increase the minimum support or reduce approximation factors. On DBpedia the number of path patterns does not grow when~$k>2$, so run time does not increase either (see Table~\ref{table:numberpathspatternsinrules}). This is because some vertices have no outgoing edges, thus most path patterns have length~2. In Pokec, when the $k>2$, the number of reachability patterns increases, thus the run time increases.
Overall, our approximation methods work well when~$k$ is large, as the run time gap between the exact and approximate algorithms widens.

\input{tables/numbers}

\smallskip\noindent
{\bf Varying approximation factors~$\psi$ and~$\rho$}. The approximation factors~$\psi$ for candidate reduction and~$\rho$ for sampling indicate the degree of approximation. As both decrease, the extent of approximation increases. Figure~\ref{fig:runtime_approxiamtiohn} plots run time varying approximation factor value for both~$\psi$ and~$\rho$. Notably, when approximation factors are small, run time is low, with the partial exception of Nell. On Nell, the run time of \textsf{\method w/ SA} is high when~$\rho = 0.2$ due to many false positives that burden the rule discovery step, as Nell is a small graph compared with DBpedia and Pokec. On the other hand, candidate reduction does not increase the number of found frequent path patterns and rules, so the run time of \textsf{\method w/ CR} grows with the approximation factor. In conclusion, sampling effectively reduces run time in large graphs, yet it may not be effective on small graphs.

\subsection{CSM-A setting}\label{app:csm-a}

We juxtapose the run time of our algorithms to that of CSM-A. CSM-A has three parameters: minimum support~$\minsup$, distance threshold~$h$, and the size of outputs~$k$. We set each value in each dataset as follows: First, we set~$k$ to the number of rules found by our algorithm, as in Table~\ref{table:numberpatterns}; then we set~$h$ to zero, i.e., the lightest parameter on CSM-A. The search space grows with~$h$. Lastly, we set~$\minsup$ to a value with which CSM-A finds at least~$k$ patterns. CSM-A works on labeled graphs rather than attributed graphs, so we use one of the attributes in each vertex as a vertex label; we thus
reduce the number of vertex attributes in Nell, DBpedia, and Pokec to~1 for CSM-A, while we let our algorithms work on a larger search space than CSM-A. MiCo is a labeled graph, so our algorithms and CSM-A operate on the same graph. As CSM-A runs on a single thread, we also run our algorithms single-threaded in this comparison. We summarize the parameters in Table~\ref{table:CSMparameter}.




We report the runtime and the number of patterns to show that we set~$\minsup$ appropriately in CSM-A. Table~\ref{table:CSMminsuppot} shows the performance of CSM-A in each dataset for two values of minimum support. One minimum support outputs~$k$ rules and another minimum support outputs fewer rules than$k$. The minimum support and the number of found rules differ across methods. For example, in Nell, with our method we set the minimum support to~1,000 to find~1,373,514 rules, while CSM-A needs a minimum support of~30 to find~1,373,514 rules.

\begin{table}[!h]
\vspace{-1mm}
\caption{The parameters in CSM-A.}\label{table:CSMparameter}
\vspace{-2mm}
\scalebox{0.85}{
\begin{tabular}{crrrr}
\hline
Parameter          & \multicolumn{1}{c}{Nell} & \multicolumn{1}{c}{MiCo} & \multicolumn{1}{c}{DBpedia} & \multicolumn{1}{c}{Pokec}\\ \hline
Output size $k$ &1,373,707 & 45 & 36 & 61 \\
Distance threshold $h$& 0& 0& 0& 0\\
Minimum support $\minsup$ &30&7,000&10,000&80,000\\
\hline
\end{tabular}
}
\vspace{-1mm}
\end{table}

\begin{table}[!h]
\vspace{-1mm}
\caption{The impact of minimum support $\minsup$ in CSM-A.}\label{table:CSMminsuppot}
\vspace{-2mm}
\scalebox{0.85}{
\begin{tabular}{crrr}
\hline
Dataset & \multicolumn{1}{c}{$\minsup$}  & \multicolumn{1}{c}{Run time [sec]} & \multicolumn{1}{c}{\# patterns}\\ \hline
\multirow{2}{*}{Nell} & 30 & 835.9 & 1,373,514 \\
& 40 & 296.7 & 355,467 \\\hline
\multirow{2}{*}{Mico} & 7,000 & 228.1 & 45 \\
& 8,000 & 194.6 & 21 \\\hline
\multirow{2}{*}{DBpedia} & 10,000 & 1\,593.2 & 35 \\
& 20,000 & 821.5 & 1 \\\hline
\multirow{2}{*}{Pokec} & 80,000 & 848\,899 & 61 \\
& 90,000 & 687\,949 & 28 \\\hline
\end{tabular}
}
\vspace{-1mm}
\end{table}

\newpage
\subsection{Memory usage on synthetic graphs}\label{app:memory}

Figure~\ref{fig:memory_graphsize} shows the memory usage on synthetic graphs.
This results show the our approximation can reduce the memory usage; In particular, sampling can reduce the memory usage compared to the candidate reduction in the exponential, though the candidate reduction can accelerate run time.
This is because for large scale graphs, the sampling can significantly reduce the number of vertices to store as the source of paths.

\input{figures/memory_graph_size}

%% file: tables/notations.tex
\begin{table}[!h]
\vspace{-2mm}
\caption{Notations employed.}\label{table:notation}
\vspace{-2mm}
\scalebox{0.85}{
\begin{tabular}{cl}
\hline
Notation           & Definition\\ \hline
$\attributes$, $a$ & Set of all attributes, an attribute\\
$A(v)$             & Set of attributes associated with vertex $v$\\
$p$                & Path pattern\\
$\vertices(p)$     & Set of vertices matching path pattern~$p$\\
$A_i$              & $i^\mathrm{th}$ set of attributes in path pattern\\
$e_i$              & $i^\mathrm{th}$ edge label in path pattern\\
$r$ or $p_x \Rightarrow p_y$ & Rule, antecedent~$p_x$ and consequent~$p_y$\\
$\pathpatterns_i$  & Set of simple path patterns with $i$ path length\\
$\pathpatterns^*$  & Set of reachability path patterns\\
$\candidates$      & Set of candidates\\
$\rules$           & Set of rules\\
$\elabels_T$       & Set of target edge labels\\
$\attributes_T$    & Set of target attributes\\
$k$                & Maximum path length\\
$\theta$           & Support threshold\\
$d_m$              & The largest in-degree among vertices\\
$\edges(A, \elabel)$    & Set of~$\elabel$-labeled edges adjacent to~$A$-covering vertex\\
$\vertices(A, \elabel)$ & Set of~$A$-covering vertices with~$\elabel$-labeled outgoing edge\\
\hline
\end{tabular}}
\vspace{-2mm}
\end{table}

%% file: algorithms/discovery.tex
\begin{algorithm}[!h] 
	\caption{\method algorithm}	\label{alg:discovery}
		\DontPrintSemicolon
			    \SetKwInOut{Input}{input}
	            \SetKwInOut{Output}{output}
	             \SetKwFunction{Expand}{Expand}
	            \Input{Graph $\graph$, minimum support $\minsup$, maximum length $\minlen$}
	            \Output{set of rules \rules}
	            {\bf If} minimum support is relative {\bf then} $\minsup \gets \minsup |\vertices|$\\
              ${\pathpatterns}_0,\ldots, {\pathpatterns}_k, {\pathpatterns}^*, \rules \gets \varnothing$\\
              ${\mathcal C^A} \gets \forall a \in \attributes$\\
              \While{${\mathcal C}^A \neq \varnothing$}{
                  ${\pathpatterns}_0 \gets {\pathpatterns}_0 \cup$ {\sf DiscoverAtt(${\mathcal C}^A, \graph, \minsup$)}\\
                  ${\mathcal C}^A \gets$ {\sf GenerateCandidate($\pathpatterns_0$)}\\
              }
                $(\elabels_T,\attributes_T) \gets$ {\sf MatchedEdges($\pathpatterns_0, \graph$)}\\
            \For{$i = 1$ to $k$}{
                  ${\mathcal C}^S \gets ${\sf VerticalExtend(${\pathpatterns}_{i-1},\elabels_T,\attributes_T$)\\       
                  \While{${\mathcal C}^S \neq \varnothing$}{
                  ${\pathpatterns}_i \gets {\pathpatterns}_i \cup$ {\sf DiscoverSimple(${\mathcal C}^S, \graph, \minsup$)}\\
                  ${\mathcal C}^S \gets$ {\sf HorizontalExtend(${\pathpatterns}_i$)}\\
                    }
                }   
            }
                
            $\vertices_T \gets${\sf BFS($\elabels_T, \graph, \minlen$)}\\
            ${\mathcal C}^* \gets \forall p$ computed by $\elabels_T, \attributes_T, \pathpatterns_0$\\
              \While{${\mathcal C}^* \neq \varnothing$}{
                  ${\pathpatterns}^* \gets {\pathpatterns}^* \cup$ {\sf DiscoverReachable(${\mathcal C}^*, \graph, \minsup, \vertices_T$)}\\
                  ${\mathcal C}^* \gets$ {\sf HorizontalExtend(${\pathpatterns}^*$)}\\
                }
                
             ${\mathcal C}^R \gets \forall (p \Rightarrow p')$ such that $p,p'$ are unit path patterns $\in \pathpatterns_1 \cup \pathpatterns^*$\\
                \While{${\mathcal C}^R \neq \varnothing$}{
                  $\rules \gets \rules \cup$ {\sf DiscoverRule(${\mathcal C}^R, \graph, \minsup$)}\\
                  ${\mathcal C}^R \gets \forall (p \Rightarrow p')$ so that either $p$ or $p'$ are horizontally and vertically extended path patterns in $\rules$.\\
              }
              {\sf ComputeMetrics}$(\rules)$\\
              {\bf return} the set of rules;\\
\end{algorithm}

%% file: figures/runtime_ratio.tex
\begin{figure}[!h]
\vspace{-1mm}
    \begin{minipage}[t]{1.0\linewidth}
    \includegraphics[width=1.0\linewidth]{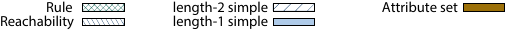}
    \end{minipage}
    \begin{minipage}[t]{1.0\linewidth}
    \centering
    \includegraphics[width=0.8\linewidth]{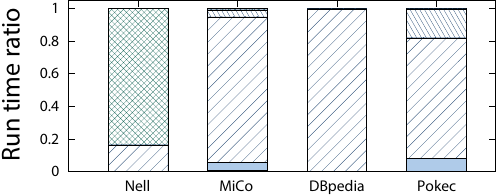}
\vspace{-3mm}    
\caption{Run time ratio}\label{fig:runtime_ratio}
\vspace{-2mm}
\end{minipage}
\end{figure}

%% file: figures/runtime_vs_length.tex
\begin{figure*}[!t]
  \subfloat[Nell]{\epsfig{file=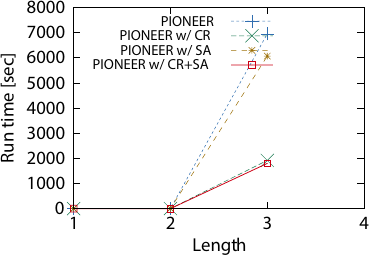,width=0.25\linewidth}
  }
  \subfloat[MiCo]{\epsfig{file=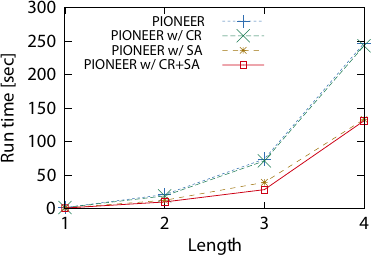,width=0.25\linewidth}
  }
  \subfloat[DBpedia]{\epsfig{file=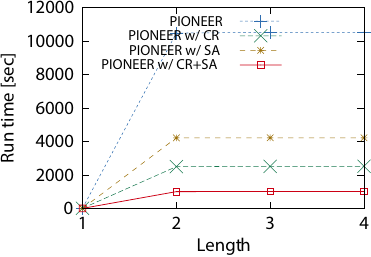,width=0.25\linewidth}
  }
  \subfloat[Pokec]{\epsfig{file=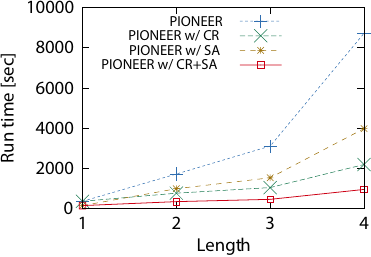,width=0.25\linewidth}
  }
\vspace{-3mm}
\caption{Impact of length on run time; missing plots indicate that the methods did not finish within 24 hours.}\label{fig:runtime_length}
\vspace{-1mm}
\end{figure*}

%% file: figures/runtime_vs_approxfactors.tex
\begin{figure*}[!t]
  \subfloat[Nell]{\epsfig{file=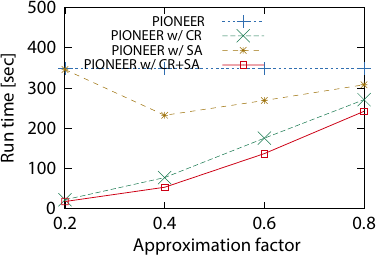,width=0.25\linewidth}
  }
  \subfloat[MiCo]{\epsfig{file=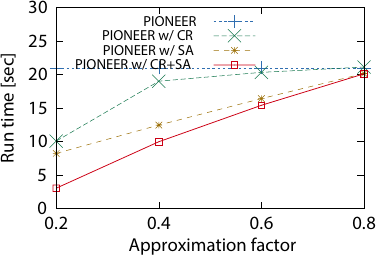,width=0.25\linewidth}
  }
  \subfloat[DBpedia]{\epsfig{file=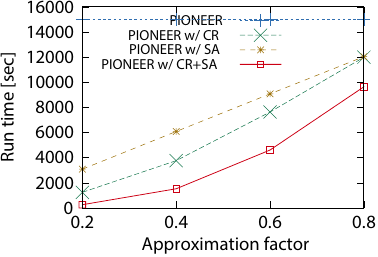,width=0.25\linewidth}
  }
  \subfloat[Pokec]{\epsfig{file=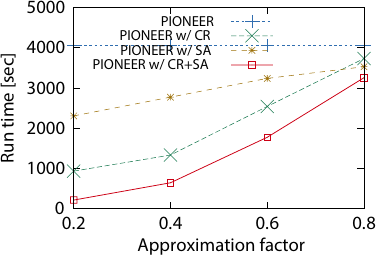,width=0.25\linewidth}
  }
\vspace{-3mm}
\caption{Impact of approximation factors on run time}\label{fig:runtime_approxiamtiohn}
\vspace{-1mm}
\end{figure*}

%% file: tables/numbers.tex
\begin{table}[!h]
\vspace{-1mm}
\caption{Numbers of path patterns and rules.}\label{table:numberpatterns}
\vspace{-3mm}
\scalebox{1}{
\begin{tabular}{l|rrrr}
\hline
& \multicolumn{1}{c}{Nell} &\multicolumn{1}{c}{MiCo}& \multicolumn{1}{c}{DBpedia} & \multicolumn{1}{c}{Pokec} \\ \hline
Attribute sets & 21 & 7& 14 & 8\\
Simple paths & 2,275& 18 & 6  & 10 \\
Reachability paths & 35& 6 & 6 & 31 \\
Rules       & 1,373,707& 45 & 36 & 61 \\
\hline
\end{tabular}}
\vspace{-3mm}
\end{table}

\begin{table}[!h]
\caption{Numbers of pattern types in rules.}\label{table:numberpathspatternsinrules}
\vspace{-3mm}
\scalebox{1}{
\begin{tabular}{l|rrrr}
\hline
& \multicolumn{1}{c}{Nell} &\multicolumn{1}{c}{MiCo}& \multicolumn{1}{c}{DBpedia} & \multicolumn{1}{c}{Pokec} \\ \hline
1 simple paths & 8,828 & 16& 36 & 20\\
2 simple paths & 2,708,882 & 55 & 0  & 0 \\
Reachability paths & 15,704 & 19 & 36 & 102 \\
\hline
\end{tabular}}
\vspace{-1mm}
\end{table}

%% file: figures/memory_graph_size.tex
\begin{figure}[!h]
\vspace{-1mm}
  \subfloat[Uniform]{\epsfig{file=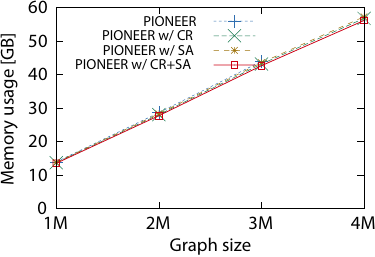,width=0.5\linewidth}
  }
\subfloat[Exponential]{\epsfig{file=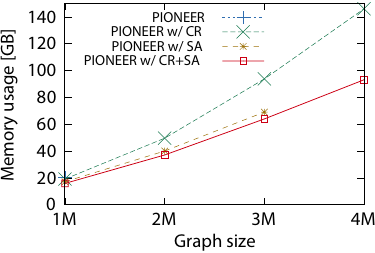,width=0.5\linewidth}
  }
\vspace{-3mm}
\caption{Memory usage to graph size.}\label{fig:memory_graphsize}
\vspace{-2mm}
\end{figure}